\documentclass[preprint,showpacs,aps]{revtex4}

\usepackage{epsfig}
\usepackage{color}

\begin{document}
\title{\large \bf Nonsteady dynamics properties of a domain wall for the creep state under an alternating driving field}

\author{N.J. Zhou$^{1}$\footnote{corresponding author; email: zhounengji@hznu.edu.cn}, and B. Zheng$^{2}$}

\affiliation{$^{1}$Department of Physics, Hangzhou Normal University, Hangzhou 310036, P.R. China \\
$^{2} $Department of Physics, Zhejiang University, Hangzhou 310027, P.R. China}

\begin{abstract}
With Monte Carlo simulations, the nonsteady dynamics properties of a domain wall have been systematically investigated
for the thermally activated creep state under an alternating driving field.
Taking the driven random-field Ising model in two dimensions as an example, two distinct growth stages
of the domain interface are identified with both the correlation length and roughness function. One stage
belongs to the universality class of the random depositions,
and the other to that of the quenched Edwards-Wilkinson equation.
In the latter case, due to the dynamic effect of overhangs, the domain interface may exhibit an intrinsic
anomalous scaling behavior, different from that of the quenched Edwards-Wilkinson equation.
\end{abstract}

\pacs{64.60.Ht, 05.10.Ln, 75.60.Ch}

\maketitle
\section{Introduction}
In recent years, the domain-wall motion has become a source of much experimental and theoretical research
\cite{tat04, kle07a, mar07, met07, kim09, sch11, mir11, lee11}.
The dynamics under an alternating driving field has attracted extensive interests in vortex lattices \cite{dol10,cao12}, liquid crystals \cite{jez08},
ferromagnetic/ferroelectric materials \cite{kag11,ste12,zhe13} and crystalline solids \cite{lau12}.
In particular, the magnetic domain-wall dynamics is an important topic in nanomaterials, thin films and semiconductors,
because of its potential technological applications including magnetic random access memories and logic devices \cite{yam07,hay08,par08}.
In the experiments of ultrathin ferromagnetic and ferroelectric films,
considerable attention is devoted to the complex susceptibility
($\chi=\chi' - i\chi''$) \cite{ven13,bra05,kle07}, which depicts the domain-wall motion.
Four dynamic states are observed in the Cole-Cole diagram of $\chi'$ vs$.$ $\chi''$,
which are relaxation, creep, sliding and switching.
Recently, experimental evidences of the relaxation-to-creep dynamic transition have been found, not only in
ultrathin ferromagnetic trilayers and ferroelectric films \cite{bra05,kle07,kle06a,yan10},
but also in liquid crystals, ferroelastic materials and molecular ferrimagnets \cite{har04, jez08, mus11}.
However, theoretical understanding of the transition is limited, especially
for the growth process of the correlation length \cite{nat01,gla03}.

For the creep state at low temperatures and low frequencies, one observes an inverse power-law behavior
for the complex susceptibility $\chi(f)=\chi_{\infty}[1+ (i2\pi f\tau)^{-\beta}]$ \cite{kle07a}.
Here $\chi_{\infty}$ denotes the bulk background susceptibility when the frequency $f \rightarrow \infty$, $\tau$ is the characteristic relaxation time,
and $\beta$ is the creep exponent. According to scaling arguments, a theoretical value
$\beta = (2 - 2\zeta)/z \approx 0.5$ is expected with the roughness exponent
$\zeta \approx 2/3$ and dynamic exponent $z \approx 1.33$ \cite{kle07}. While experimental values of $\beta$
vary from $0.2$ to $0.65$ in ultrathin ferromagnetic and ferroelectric films \cite{bra05, kle06a, kle07}.
Hence, it remains much challenging to understand the creep exponent.

Up to date, theoretical tools for describing the domain-wall motion are typically based on the Edwards-Wilkinson equation with
quenched disorder (QEW) \cite{pet04,kol06,due05,fer13,sch11,sch10}. With this equation, the dynamics properties for the creep state
under a constant driving field or zero field have been well understood \cite{kol05,kol09,cha00}. It can be viewed as a thermally activated hopping movement from
one local energy minimum to the next, dominated by the energy barrier $U_B$ that must be overcome.
The energy barrier grows as a power law $U_B(\xi) \sim \xi^{\psi}$, responsible for the logarithmical
growth of the correlation length $\xi(t) \sim (\ln t)^{1/\psi}$ \cite{mon08,mon08b}.
An effective energy barrier exponent $\psi \approx 0.49$ has been numerically measured \cite{kol05a,mon08a},
and the roughness exponent $\zeta \approx 2/3$ has also been estimated from the kinetic roughening of the domain wall \cite{kol05,kol09}.
However, few works deal with the creep dynamics under an alternating driving field. Moreover,
detailed microscopic structures and interactions of real materials are not concerned
in the phenomenological QEW equation \cite{pet04}.

To further understand the creep dynamics from a more fundamental viewpoint, we should build lattice models
which allow a closer comparison between theory and experiment. The driven random-field Ising model (DRFIM) is a candidate,
which has been used to understand the dynamic transitions in ferroic systems \cite{col06,zho09,zho10}.
Despite not including all interactions in real materials, it at least captures robust features of the domain-wall motion.
Very recently, the creep motion of a domain wall driven by a constant field has been numerically investigated
with the DRFIM model, and the results are comparable with experiments \cite{don12a}.

Taking the two-dimensional ($2D$) DRFIM model as an example, we conduct a comprehensive study on the creep dynamics
under an alternating driving field. With Monte Carlo simulations,
we accurately determine the scaling exponents $\beta, \psi$ and $\zeta$,
and identify the universality classes, in comparison with those of the QEW equation and experiments.
In Sec. II, the model and scaling analysis are described, and in Sec. III, the numerical results are
presented. Finally, Sec. IV includes the conclusions.

\section{model and scaling analysis}
The DRFIM model is defined by the following Hamiltonian
\begin{equation}
\mathcal{H} = - J \sum_{\langle ij\rangle}S_iS_j - \sum_i
\left[h_i+H(t)\right] S_i, \label{equ100}
\end{equation}
where $S_i = \pm 1$ is a Ising spin at site $i$ of the lattice,
$\langle ij \rangle$ denotes a nearest-neighbor pair of spins, and
$h_i$ is a quenched random field uniformly distributed within an
interval $[-\Delta, \Delta]$. We use a homogeneous
alternating driving field $H(t)=H_0\cos(2\pi f t)$, and set the coupling constant $J=1$ \cite{zho10}.
In order to make sure that the dynamic evolution of spins occurs at or around
the domain wall, we restrict the temperature $T \leq 0.66$, the
disorder strength $\Delta \leq 2.0$ and the driving field $H_0
\leq 0.5$. Simulations are performed on a rectangular lattice $L_x
\times L_y$ with the antiperiodic and periodic boundary conditions
along the $x$ and $y$ directions, respectively.

The initial state is a {\it semiordered} state with a perfect domain
wall in the $y$ direction.  To eliminate the pinning effect
irrelevant for the disorder, we rotate the lattice such
that the initial domain wall orients in the $(11)$ direction \cite{zho09,zho10a}. After
preparing the initial state, we update spins with the {\it heat-bath
algorithm} \cite{zhe98}. As time evolves, the domain wall moves and roughens,
while the bulk remains unchanged. Therefore, the domain wall can also be called
{\it a domain interface} \cite{zho07,zho08}. Main results of numerical simulations are
presented with the lattice size $L_x=25$ and $L_y=512$, up to
$t_{max}=400~000$ Monte Carlo steps (MCS). Here MCS is defined by
$L_x \times L_y$ single-spin-flips attempts. For each set of model
parameters ($T, \Delta, H_0, f$), more than $10~000$
samples are performed for average. Errors are estimated by dividing
the samples into three subgroups. If the fluctuation of the
curve in the time direction is comparable with or larger than the
statistical error, it will also be taken into account. Additional
simulations with $L_x=50$ are performed to confirm that the
finite-size effect is negligible.

Denoting a spin at site $(x,y)$ by $S_{xy}(t)$, we first introduce
the height function
\begin{equation}
h(y,t) = \sum_{x=1}^{L_x} S_{xy}(t),
\label{equ110}
\end{equation}
and then define the position of the domain interface
\begin{equation}
h(t) = \frac{1}{2}\left[ \left \langle h(y,t) \right \rangle +
L_x \right ].
\label{equ120}
\end{equation}
Here $\langle \cdots \rangle$ represents not only the statistic
average over Monte Carlo samples, but also the average in the $y$
direction. After the stationary magnetic hysteresis loop is obtained at $t>t_0$,
the complex susceptibility can be calculated by \cite{pet04,zho10},
\begin{equation}
\chi(f, T) = \frac{1}{PH_0}\int_0^{P}h(t)e^{-i2\pi ft} dt,
\label{equ130}
\end{equation}
where $P= 1/f$ is the time period of the alternating driving field.

With the height function $h(y,t)$ at hand, the roughness function
$\omega^{2}(t)$ and the correlation function $C(r,t)$ are defined respectively by
\begin{eqnarray}
\omega^{2}(t) & = & \left \langle h(y,t)^2 \right \rangle - \left \langle h(y,t) \right \rangle^2
\label{equ140}
\end{eqnarray}
and
\begin{eqnarray}
C(r, t) & =& \left\langle h(y,t)h(y+r,t) \right \rangle - \left
\langle h(y,t) \right \rangle^2.
\label{equ145}
\end{eqnarray}
$\omega^{2}(t)$ and $C(r,t)$ describe the roughening of the domain interface in the
$x$ direction and the growth of the spatial correlation in the $y$
direction, respectively. To reveal the characteristics of the
thermally activated creep dynamics, we introduce the creep susceptibility
\begin{equation}
D\chi'(f, T)  =  \chi'(f, T) - \chi'(f, T=0),
\label{equ150}
\end{equation}
and the pure roughness function
\begin{equation}
D\omega^{2}(t) = \omega^{2}(t) - \omega^{2}(t, T=0).
\label{equ160}
\end{equation}
To detect overhangs generated in the
creep motion, another two definitions of the height
functions, $h^{+}(t)$ and $h^{-}(t)$, are introduced by the envelopes of
the positive and negative spins, respectively \cite{zho10a}. It is believed that
the difference $Dh(t)=h^{+}(t) - h^{-}(t)$ describes the
average size of overhangs.

According to the phenomenological scaling arguments \cite{fed05}, a power-law dispersion of the creep
susceptibility is obtained for the creep dynamics under an alternating driving field,
\begin{equation}
D\chi'(f) \sim (1/f)^{\beta}.
\label{equ170}
\end{equation}
For the $\xi$-length domain-wall segments, a certain hopping time $t \sim \exp(U_B(\xi)/T)$ is required
to overcome the energy barrier $U_B(\xi)$ \cite{kol05a}. Assuming that the energy barrier
scales as $U_B(\xi) \sim \xi^{\psi}$, one may deduce
\begin{equation}
D\xi(t) \sim [T\ln(t)]^{1/\psi}.
\label{equ180}
\end{equation}
Here $D\xi(t) = \xi(t) - \xi(t, T=0)$ is the so-called creep
correlation length, and $\psi$ is the energy barrier exponent.

For a sufficiently large lattice $L \gg \xi(t)$, the dynamic
behavior of $\xi(t)$ can be extracted from
the correlation function \cite{kol06, zho13},
\begin{equation}
C(r,t) = \omega^{2}(t) \widetilde{C}\left( r/\xi(t) \right),
\label{equ190}
\end{equation}
where $\widetilde{C}(s)$ is the scaling function with $s=r/\xi(t)$,
and $\omega^{2}(t)$ is the roughness function defined in
Eq.~(\ref{equ140}). In the kinetic roughening of the domain
interface, a power-law scaling behavior of the pure
roughness function is expected with the roughness exponent $\zeta$,
\begin{equation}
D\omega^{2}(t) \sim \left[D\xi(t)\right]^{2\zeta}.
\label{equ200}
\end{equation}
Meanwhile, one may determine the local roughness
exponent $\zeta_{loc}$ by fitting $C(r, t)$ with an empirical scaling form \cite{jos96},
\begin{equation}
C(r, t) = A \left[\tanh\left( r / \xi(t)\right) \right]^{2\zeta_{loc}}.
\label{equ205}
\end{equation}

\section{Monte Carlo simulations}
\subsection{Numerical results}
\begin{figure}[ht]
\epsfysize=9.5cm \epsfclipoff \fboxsep=0pt
\setlength{\unitlength}{1.cm}
\begin{picture}(10,6)(0,0)
\put(0.5,-1.0){{\epsffile{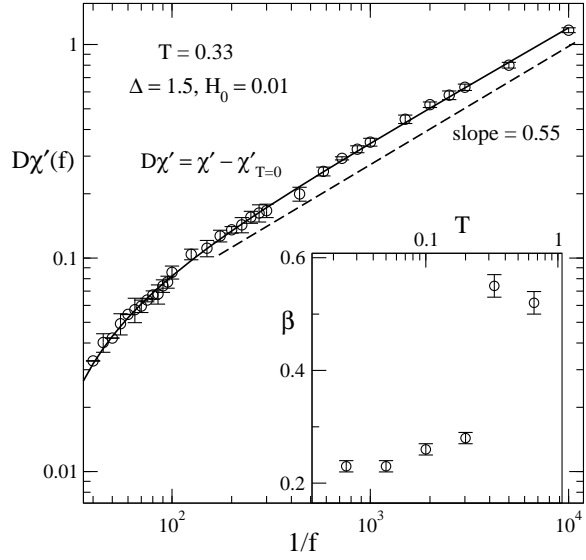}}}
\end{picture}

\caption{The spectrum of the creep susceptibility $D\chi'(f)$ is
displayed on a log-log scale, at the temperature $T=0.33$,
the strength of the disorder $\Delta= 1.5$ and the driving field $H_0=0.01$.
The dashed line represents a power-law fit,
and the solid line includes the correction in the form $y = ax^{\beta}(1 - c/x)$.
In the inset, the creep exponent $\beta$ at different $T$ is plotted.
Error bars are given when they are larger than or compatible with the symbols.
} \label{f1}
\end{figure}

In Fig.~\ref{f1}, the spectrum of the creep
susceptibility $D\chi'(f)$ defined in Eq.~(\ref{equ150}) is plotted on a log-log scale at
the temperature $T=0.33$, the strength of the disorder $\Delta = 1.5$ and the
driving field $H_0=0.01$.  To obtain stationary results, the data in a waiting time $t_0=20$ periods are skipped in
the computation of $\chi(f,T)$. A power-law behavior is observed but with certain corrections to scaling.
A direct measurement from the slope yields the exponent $\beta = 0.55(2)$, and the correction in the form
$y = ax^{\beta}(1 - c/x)$ extends the fitting to the early times but with the same value of $\beta$ within the error bar.
For comparison, the creep exponents at other temperatures $T = 0.025$, $0.05$, $0.1$, $0.2$ and $0.66$ are measured.
As shown in the inset, the result $\beta \approx 0.2$ jumps suddenly to $0.5$ around $T=0.33$.
In order to understand the underlying dynamic mechanism, we investigate the nonsteady dynamics in the following.

\begin{figure}[ht]
\epsfysize=8.5cm \epsfclipoff \fboxsep=0pt
\setlength{\unitlength}{1.cm}
\begin{picture}(10,6)(0,0)
\put(-3.0,-1.0){{\epsffile{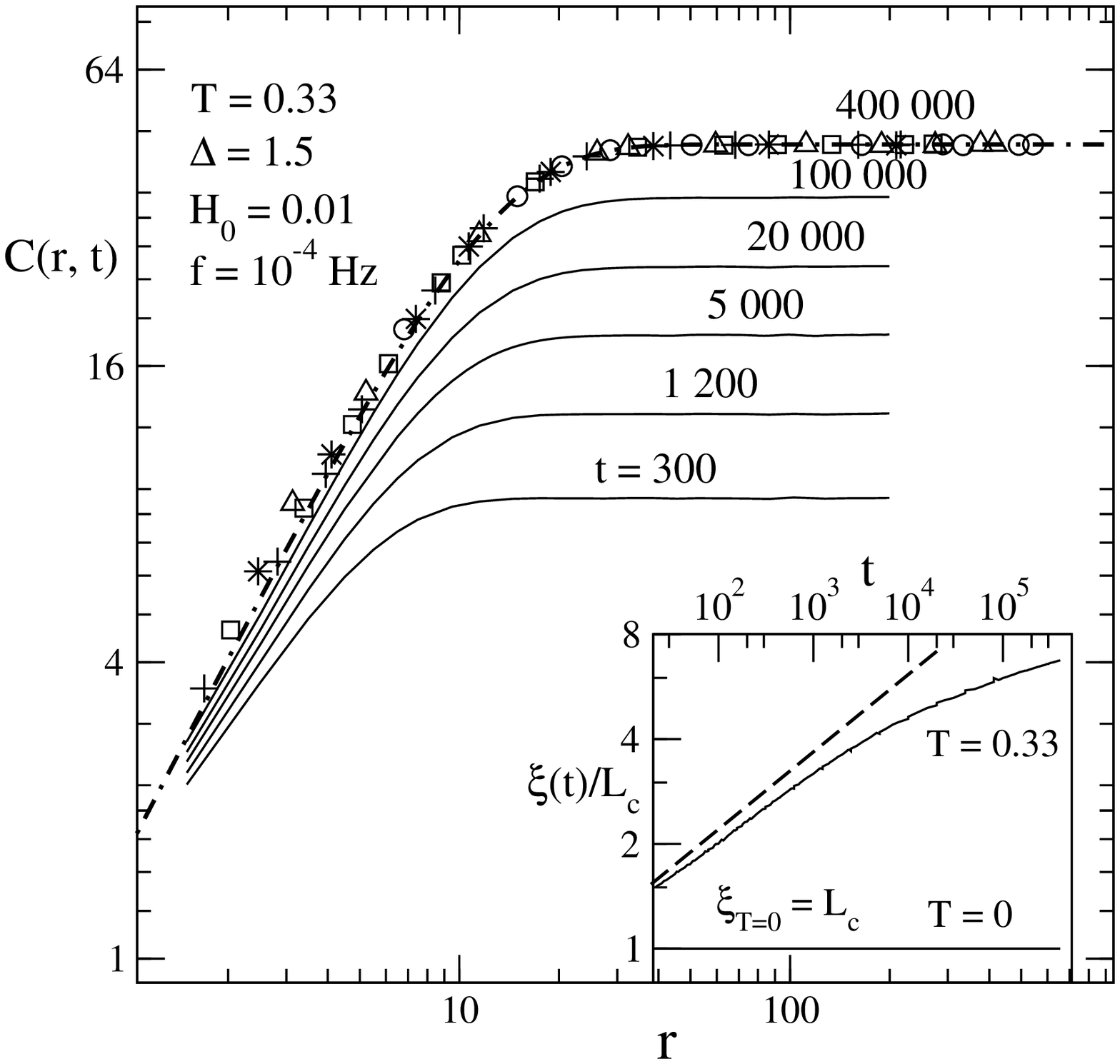}}}\epsfysize=8.5cm
\put(5.2,-1.0){{\epsffile{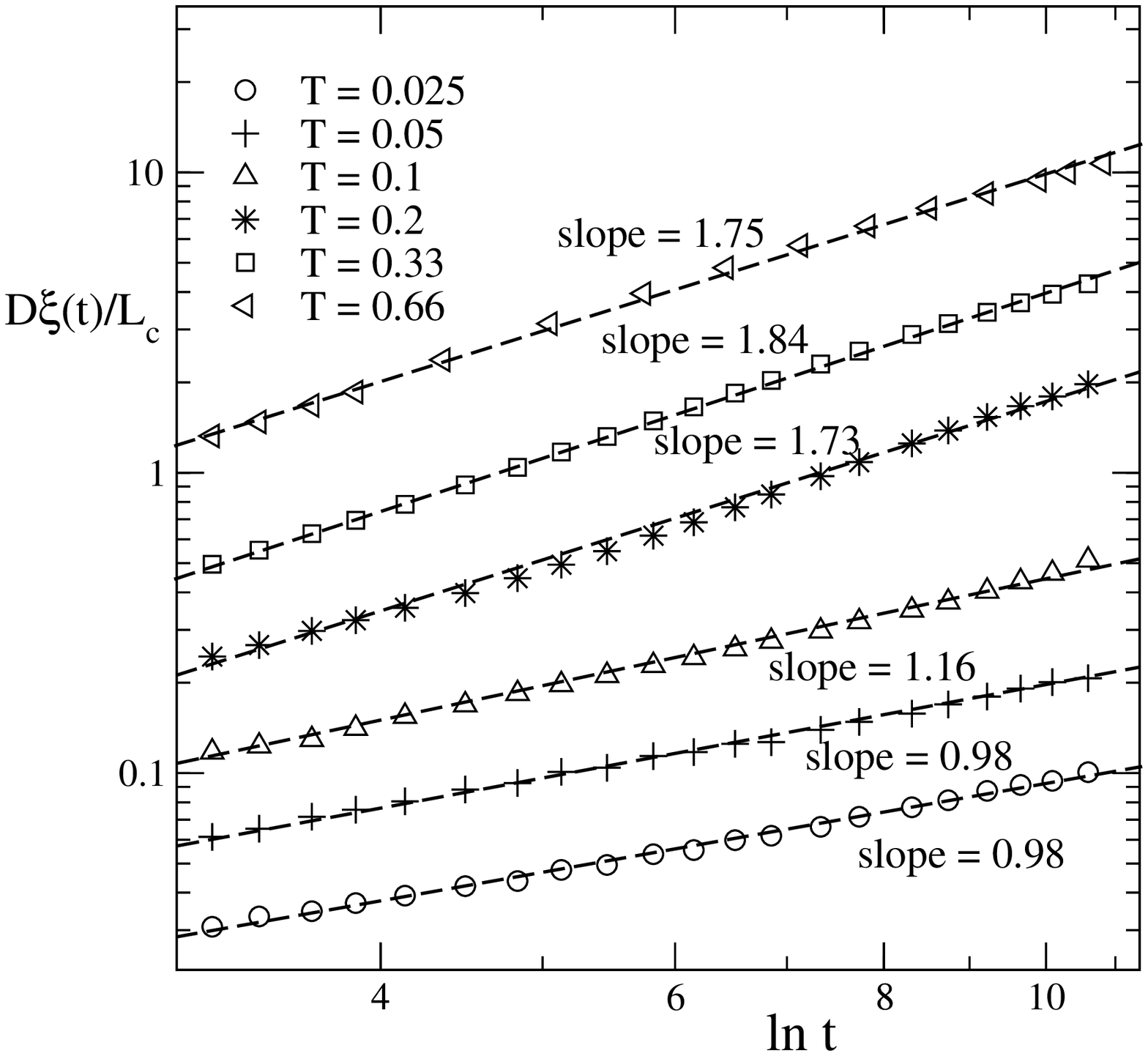}}}
\end{picture}

\hspace{1.0cm}\footnotesize{(a)}\hspace{8.0cm}\footnotesize{(b)}
\caption{(a) The correlation function $C(r, t)$ at different times $t$ are
plotted with solid lines. Symbols show data collapse, and the dash-dotted line represents
a fit according to Eq.~(\ref{equ205}). In the inset, the rescaled spatial correlation length $\xi(t)/L_c$
is plotted for $T=0$ and $0.33$ on a log-log scale. (b) The creep correlation length
$D\xi(t)=\xi(t) - \xi_{_{T=0}}$ rescaled by $L_c$ is displayed as a function of $\ln t$
for different $T$. In both (a) and (b), dashed lines show power-law fits.
}\label{f2}
\end{figure}

Taking the set of model parameters ($T=0.33, \Delta=1.5, H_0=0.01$
and $f=10^{-4}$ Hz ) as an example, the correlation function
$C(r,t)$ is displayed as a function of $r$ in Fig.~\ref{f2}(a).
According to Eq.~(\ref{equ205}), a perfect fitting to the numerical
data is observed, and the local roughness exponent $\zeta_{loc} =
0.65(1)$ is measured. Based on the scaling form of $C(r,t)$ in
Eq.~(\ref{equ190}), numerical data of different time $t$ nicely
collapse to the curve at $t'=400~000$ MCS by rescaling $r$ to
$\left[\xi(t')/\xi(t)\right]r$ and $C(r,t)$ to
$\left[\omega^2(t')/\omega^2(t)\right]C(r,t)$. With the
data-collapse technique \cite{don12a}, we extract the nonequilibrium
correlation length $\xi(t)$ from the correlation function $C(r,t)$.
In the inset, the dynamic evolution of $\xi(t)$ is displayed on a
log-log scale. The significant deviation from the power-law behavior
indicates that the correlation length $\xi(t)$ does not obey the
usual growth law $\xi(t) \sim t^{1/z}$ \cite{kol05a,don12}.
Additionally, a time-independent correlation length $\xi = L_c$ is
observed at $T=0$ for the relaxation state, irrelevant to the creep
dynamics. Therefore, we define the creep correlation length
$D\xi(t)$ by subtracting the relaxation length $L_c$, and introduce
a dimensionless correlation length $D\xi(t)/L_c$.

\begin{figure}[ht]
\epsfysize=8.5cm \epsfclipoff \fboxsep=0pt
\setlength{\unitlength}{1.cm}
\begin{picture}(10,6)(0,0)
\put(-3.0,-1.0){{\epsffile{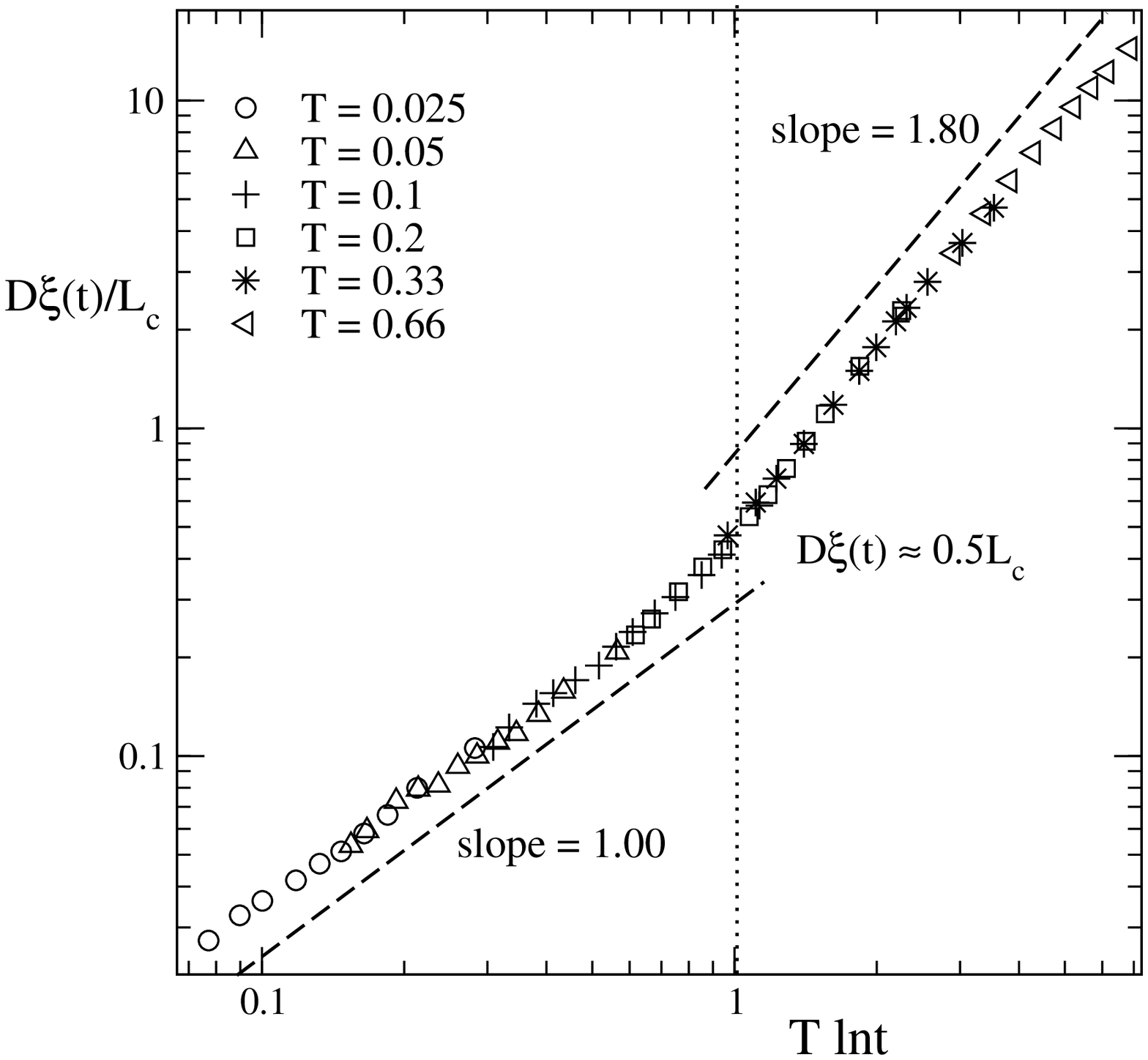}}}\epsfysize=8.5cm
\put(5.2,-1.0){{\epsffile{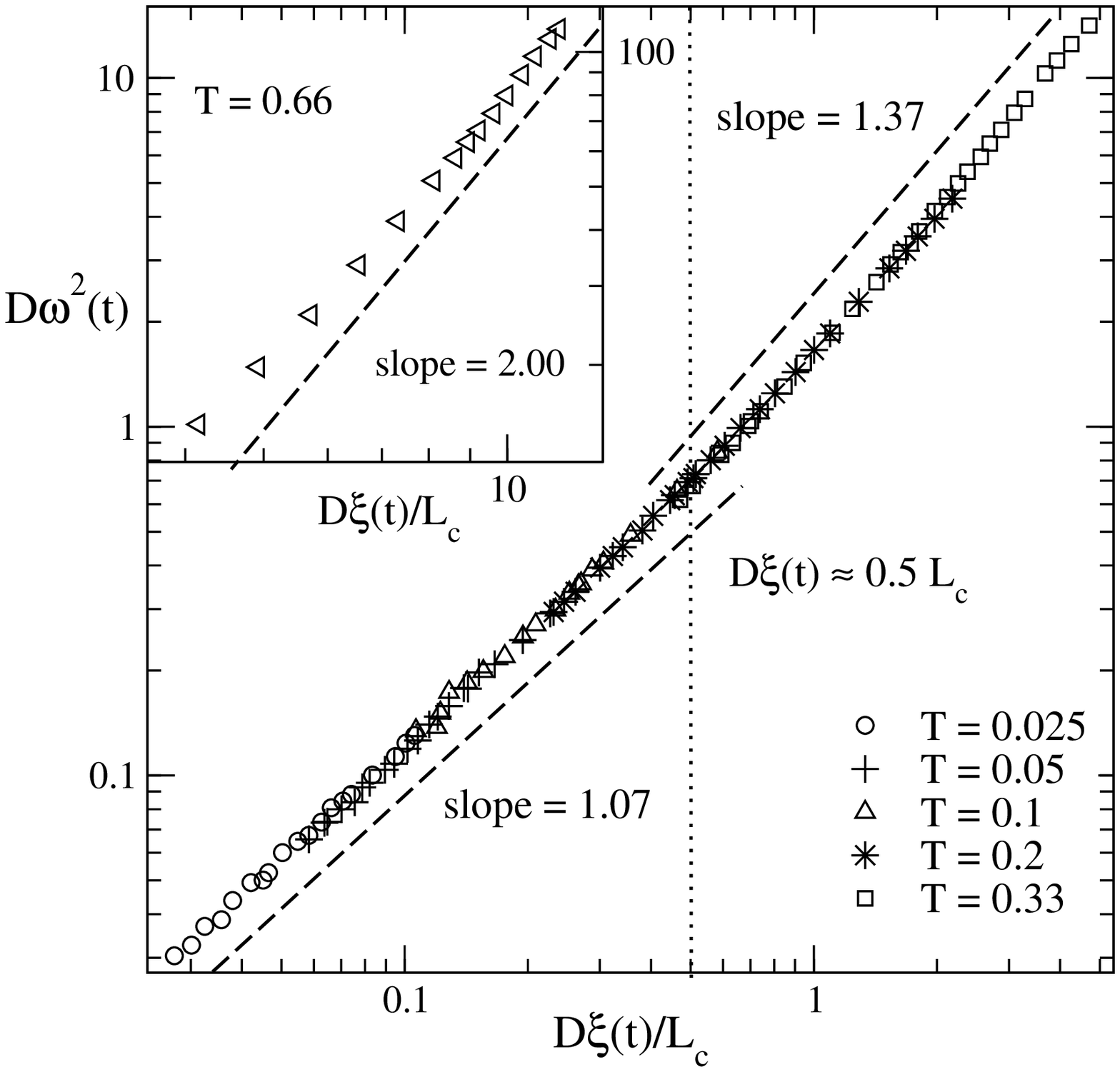}}}
\end{picture}

\hspace{1.0cm}\footnotesize{(a)}\hspace{8.0cm}\footnotesize{(b)}
\caption{(a) The creep correlation length $D\xi(t)/L_c$ vs$.$ $T\ln t$ and
(b) the pure roughness function $D\omega^2(t)$ vs$.$ $D\xi(t)/L_c$ are
plotted at different $T$. In the inset,
the asymptotic behavior of $D\omega^2(t)$ is shown at $T=0.66$.
In both (a) and (b), data collapse is demonstrated.
Dashed lines represent power-law fits, and vertical dotted lines indicate
a crossover at $D\xi(t) \approx 0.5 L_c$. }\label{f3}
\end{figure}

In Fig.~\ref{f2}(b), the creep correlation length $D\xi(t)/L_c$ is displayed
as a function of $\ln t$ at different $T$ on a log-log scale.
Power-law behaviors are observed, and the effective energy barrier exponents
$1/\psi=0.98(1),0.98(1),1.16(1),1.73(2), 1.84(2)$ and $1.75(3)$
are estimated from the slopes of the curves for
$T=0.025, 0.05, 0.1, 0.2, 0.33$ and $0.66$, respectively. To confirm the
scaling form in Eq.~(\ref{equ180}), $D\xi(t)/L_c$ vs$.$ $T \ln t$ is
plotted in Fig.~\ref{f3}(a) on a double-log scale. Data of different $T$ nicely collapse to a
master curve, and two distinct scaling regimes are detected with the slopes
$1/\psi=1.00(1)$ and $1.80(2)$. Between the two regimes, a dynamic crossover occurs
at $T\ln t \approx 1$ and $D\xi(t) \approx 0.5 L_c$.
Then we extract the characteristic of the energy barrier
\begin{equation}
U_B \sim T \ln t \sim \left\{
   \begin{array}{lll}
    2   D\xi(t)/ L_c ,     & \quad &  \mbox{ $D\xi(t) \ll 0.5 L_c$} \\
    1.5 \left(D\xi(t)/ L_c \right)^{0.56},   & \quad &  \mbox{ $D\xi(t) \gg 0.5 L_c$}
   \end{array}, \right.
\label{equ210}
\end{equation}
in the small-$D\xi(t)$ and large-$D\xi(t)$ scaling regimes, respectively.

With the creep correlation length $D\xi(t)/L_c$ at hand, we measure
the roughness exponent $\zeta$ in Eq.~(\ref{equ200}). Since the
amplitude of the alternating field $H_0 = 0.01$ is much smaller than
the depinning field $H_c=1.2933(2)$ \cite{zho09}, the roughness
exponent is actually the equilibrium exponent, though the
equilibrium state is not yet reached \cite{kol05a}.  In
Fig.~\ref{f3}(b), $D\omega^2(t)$ is plotted as a function of
$D\xi(t)/L_c$ on a log-log scale. Similarly, data collapse of
different $T$ is displayed with different symbols. In the
small-$D\xi(t)$ regime, the slope of the master curve yields the
roughness exponent $\zeta = 0.53(1)$, close to $1/2$. It suggests
that the domain interface belongs to the universality class of the
random depositions \cite{yan95,sar96,fed04,ago13}. In the
large-$D\xi(t)$ regime, $\zeta = 0.68(1)$ is estimated, in good
agreement with the equilibrium value $\zeta_{eq} = 2/3$
 of the QEW equation \cite{kol05a, mon08,kol09,igu09}. Hence, it
belongs to the universality class of the QEW equation. In the inset,
however, a noticeable increase of $\zeta$ is observed at the high
temperature $T=0.66$, and the asymptotic value is $\zeta =1.00(2)$.

\begin{figure}[ht]
\epsfysize=8.5cm \epsfclipoff \fboxsep=0pt
\setlength{\unitlength}{1.cm}
\begin{picture}(10,6)(0,0)
\put(-3.0,-1.0){{\epsffile{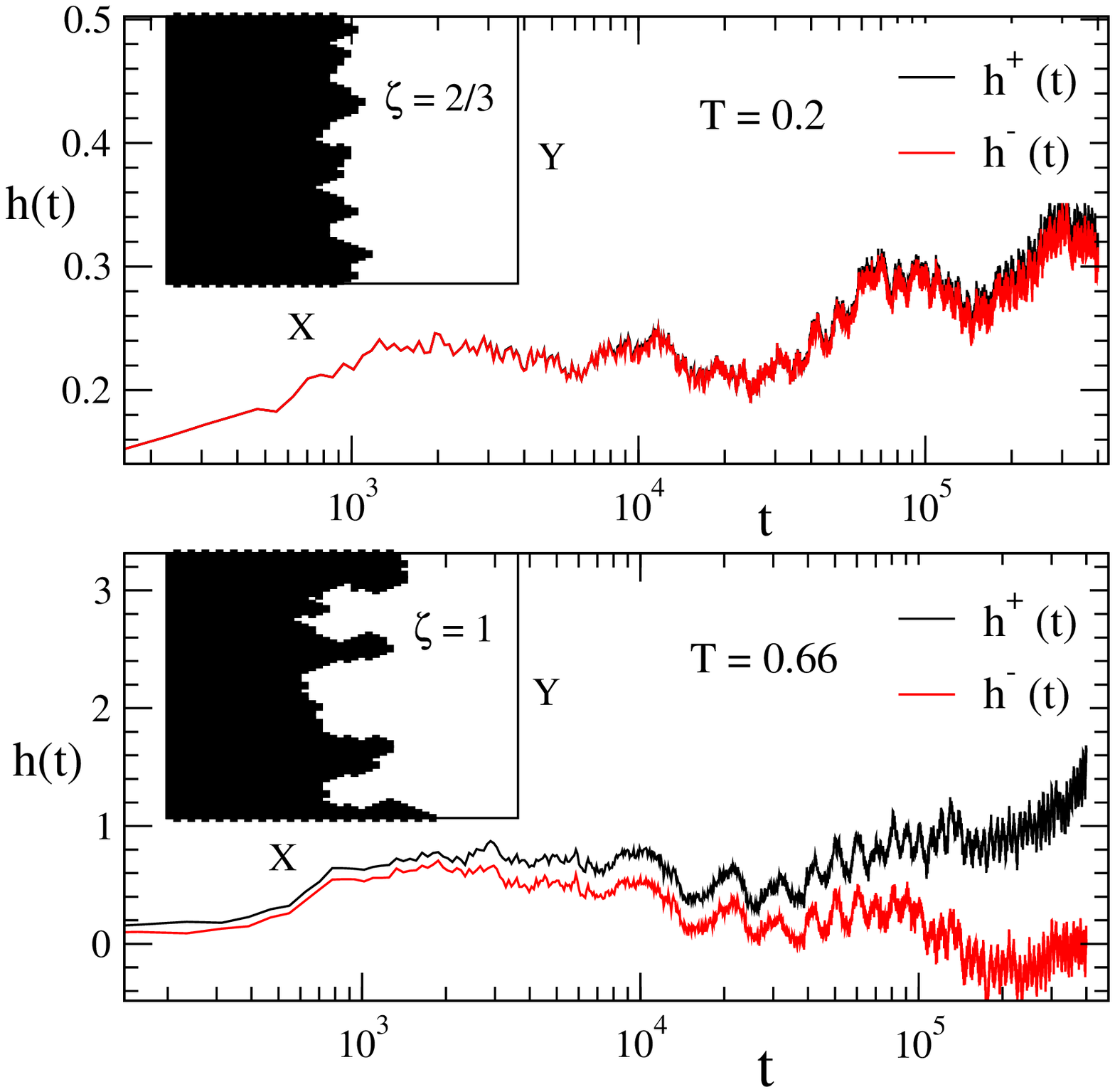}}}\epsfysize=8.5cm
\put(5.2,-1.0){{\epsffile{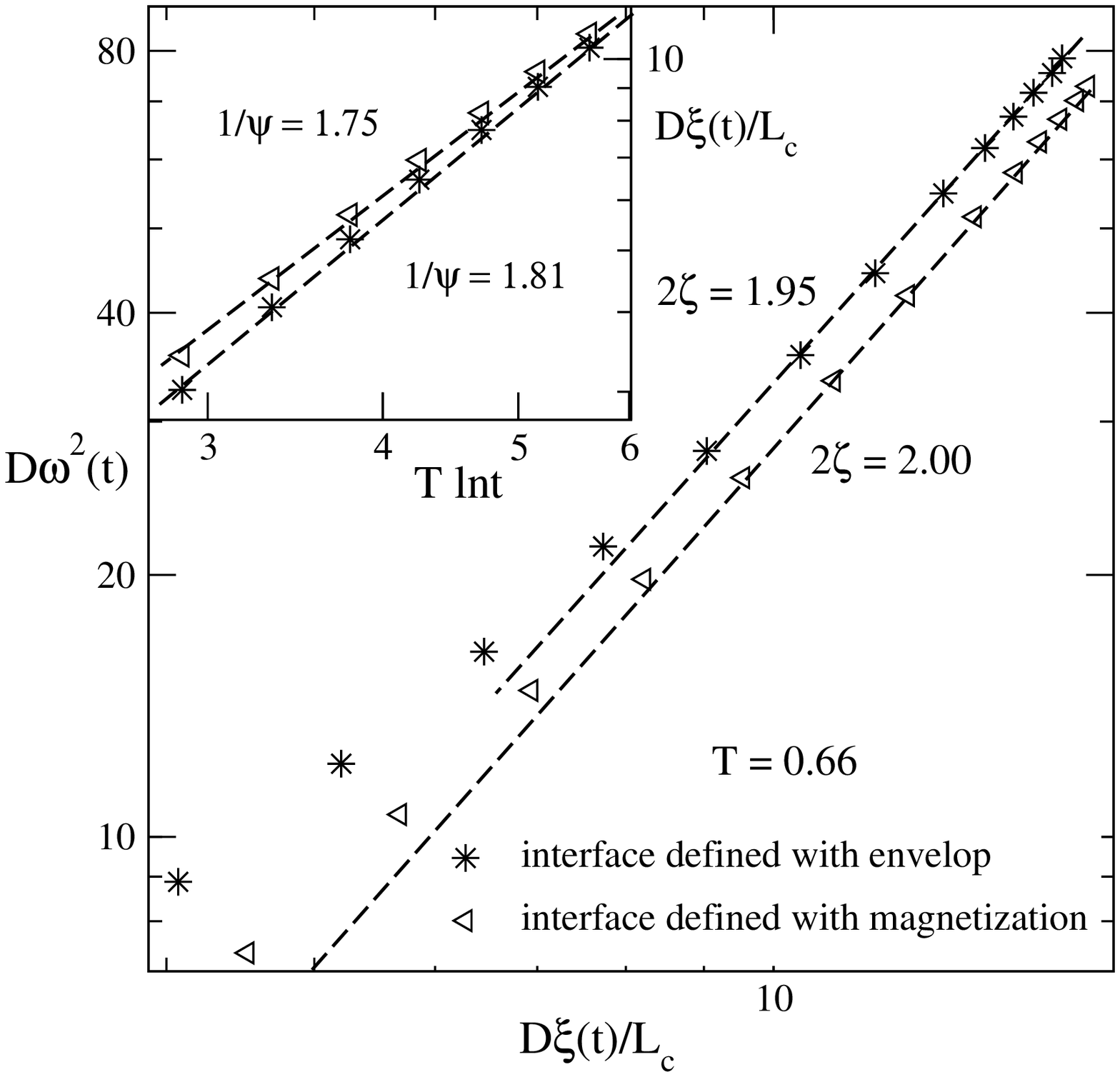}}}
\end{picture}

\hspace{1.0cm}\footnotesize{(a)}\hspace{8.0cm}\footnotesize{(b)}
\caption{ (Color on-line) (a) The time evolution of the height
functions $h^+(t)$ and $h^-(t)$ is plotted with black and red lines,
respectively. In the insets, the snapshots of the domain interfaces
at the time $t= 4 \times 10^5 $ MCS are displayed for $T=0.2$ and
$0.66$. (b) $D\omega^2(t)$ and $D\xi(t)/L_c$ at $T=0.66$ are plotted
on a log-log scale, for the domain interfaces $h(t)$ and $h^{-}(t)$
defined with the magnetization and envelop, respectively. Dashed
lines represent power-law fits. } \label{f4}
\end{figure}

To understand the unexpectedly large roughness exponent at $T=0.66$,
we examine the existence of overhangs in the creep motion
\cite{zho09, zho10a}. As shown in Fig.~\ref{f4}(a), black and red
lines represent the time evolutions of the height functions
$h^{+}(t)$ and $h^{-}(t)$, respectively. The coincidence and
noncoincidence of the curves in the upper and lower panels suggest
that the contribution of overhangs is negligible at $T=0.2$ and
important at $T=0.66$. Besides, the snapshots of the domain walls at
the time $t= 4 \times 10^5 $ MCS are also shown in the insets.
Overhangs can be observed directly in the lower panel but not in the
upper panel. Consequently, it is convincing that the overhangs
affect the dynamic evolution of the spin configuration and play an
essential role in the increase of the roughness exponent.

Due to the existence of overhangs, the position of the domain
interface $h(y,t)$ is not single-valued and the definition of the
height function is not unique. As shown in Fig.~\ref{f4}(b),
$D\omega^2(t)$ and $D\xi(t)/L_c$ at $T=0.66$ are displayed for the
domain interface $h^{-}(t)$ defined with the envelop of the negative
spins, in comparison with those for the domain interface $h(t)$
defined with the magnetization in Eq.~(\ref{equ120}). The roughness
exponent $\zeta= 0.97(2)$ and the energy barrier exponent $1/\psi =
1.81(4)$ are measured, consistent with $\zeta= 1.00(2)$ and $1/\psi
= 1.75(3)$ for the domain interface defined with the magnetization.
These results again support that the definition of the height
function with the magnetization is reasonable.

\begin{figure}[ht]
\epsfysize=8.5cm \epsfclipoff \fboxsep=0pt
\setlength{\unitlength}{1.cm}
\begin{picture}(10,6)(0,0)
\put(-3.0,-1.0){{\epsffile{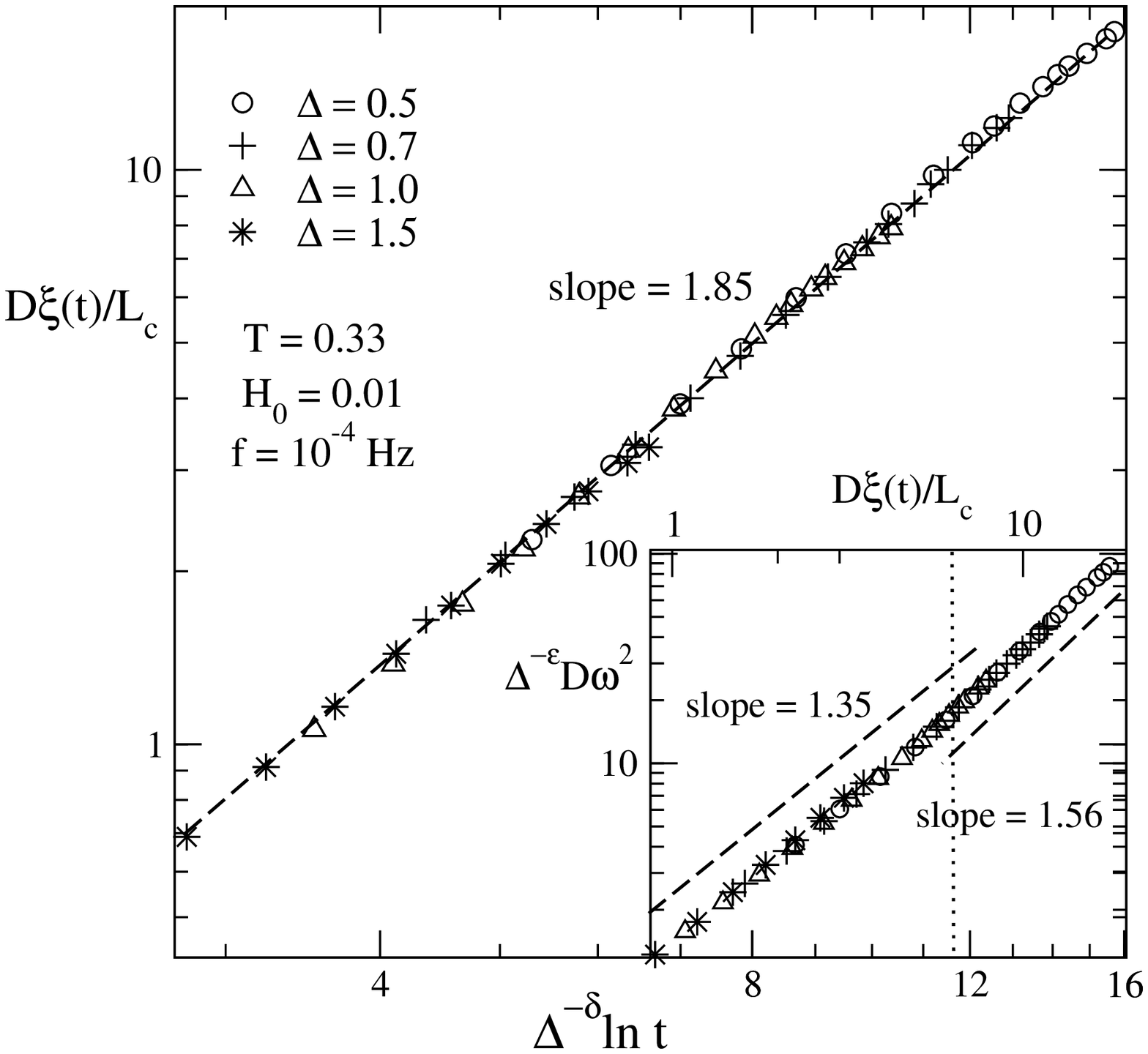}}}\epsfysize=8.5cm
\put(5.2,-1.0){{\epsffile{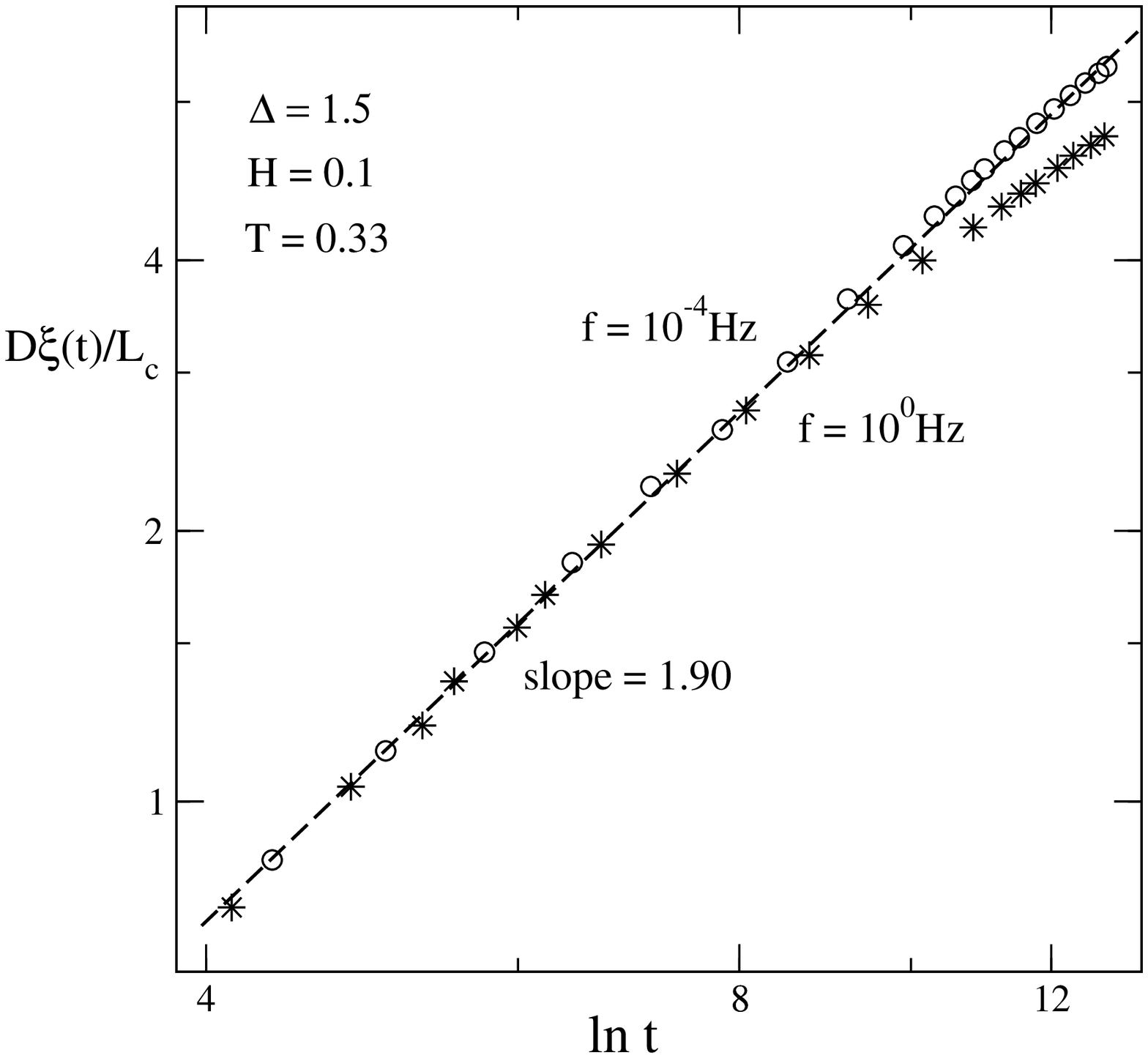}}}
\end{picture}

\hspace{1.0cm}\footnotesize{(a)}\hspace{8.0cm}\footnotesize{(b)}
\caption{(a) Taking $\delta = 0.58(1)$ as input, the creep correlation length $D\xi(t)/L_c$ is displayed against
$\Delta^{-\delta} \ln t$ for different strengthes of the disorder $\Delta$ on a log-log scale. In the
inset, $\Delta^{-\varepsilon} D\omega^2(t)$ is plotted with
$\varepsilon = 0.22(1)$ as input. The vertical dotted line indicates a crossover.
(b) A log-log plot of $D\xi(t)/L_c$ vs$.$ $\ln t$ is shown
for different frequencies $f=10^{-4}$ and $10^0$ Hz at the driving field $H_0=0.1$.
In both (a) and (b), dashed lines represent power-law fits.
}\label{f5}
\end{figure}

Besides the temperature, the effects of the quenched disorder and driving field are also investigated.
In Fig.~\ref{f5}(a), the creep correlation length $D\xi(t)/L_c$ is displayed as a function of $\Delta^{-\delta} \ln t$
at $T=0.33$ on a log-log scale. Taking
$\delta=0.58(1)$ as input, data collapse of different disorders $\Delta=0.5, 0.7, 1.0$ and $1.5$ is demonstrated,
and $1/\psi = 1.85(2)$ is determined, close to $1.80(2)$ in Fig.~\ref{f3}(a).
According to the scaling relation $\delta \approx \psi \approx 0.56$, one may derive
the scaling form of $U_B$ in the large-$D\xi(t)$ regime,
\begin{equation}
U_B \sim T \ln t \sim \left(D\xi(t) \Delta\right)^{\psi}.
\label{equ220}
\end{equation}
In the inset, the scaling function $\Delta^{-\varepsilon}D\omega^{2}(t)$ vs$.$ $D\xi(t)/L_c$
is plotted on a log-log scale.
Data of different $\Delta$ nicely collapse together with the
parameter $\varepsilon =0.22(1)$ as input. Similarly, the abnormal increase of the roughness exponent
from $\zeta = 0.68(1)$ to $ 0.78(1)$ is also induced by the dynamic effect of overhangs.

In Fig.~\ref{f5}(b), the creep dynamics of the domain wall for different frequencies $f$ is presented
at the driving field $H_0=0.1$ on a log-log scale. If the frequency is sufficiently low, e.g.,
$f=10^{-4}$ Hz, a power-law behavior of $\xi(t)$ can be observed with the exponent $1/\psi = 1.90(2)$,
somewhat larger than the one $1.80(2)$ at $H_0=0.01$. Additional simulations at $H_0=0.05,0.2$ and $0.5$
show that the energy barrier exponent $\psi$ is $H_0$-dependent, and data of different
$H_0$ are unlikely to collapse together. For a high frequency, e.g., $f=1$ Hz
corresponding to the relaxation state \cite{zho10}, $D\xi(t)$ drops obviously at the tail of the curve.
It suggests that the nonsteady dynamics properties of the relaxation state are different
from those of the creep state.

\subsection{Discussion}

The measurements of scaling exponents at different $T$ are
summarized for $H_0=0.01$ and $\Delta = 1.5$ in Table.~\ref{t1}. As
$T$ increases, the creep exponent $\beta$ changes from $0.23(1)$ to
$0.55(2)$, compatible with experimental results in the ferromagnetic
and ferroelectric films \cite{kle06a}, e.g., $\beta = 0.6(1)$ in the
ultrathin Pt/Co($0.5$nm)/Pt trilayer and $\beta = 0.35(2)$ in the
periodically poled KTiOPO$_4$ \cite{bra05,kle07}. The exponent
$\beta = 0.52(2)$ measured at the highest temperature $T=0.66$ is
consistent with the prediction of the scaling relation $\beta = (2 -
2\zeta)/z \approx 0.5$ \cite{kle07}. According to the general
scaling arguments \cite{nat90}, we propose the complex
susceptibility $D\chi(f) \sim \ln(1/f)^{\zeta/\psi} \sim
(1/f)^{k\zeta/\psi}$ with an effective coefficient $k$. Then a novel
scaling relation $\beta = k \zeta / \psi$ is obtained. As shown in
Table.~\ref{t1}, $k \approx \beta \psi / \zeta_{loc} = 0.45(3)$
holds for the whole temperature range. With this scaling relation,
one can predict the creep exponent $\beta$ by only measuring $\psi$
and $\zeta$ from the nonsteady dynamics.

Two distinct growth stages of the creep correlation length $D\xi(t)$
are found with the scaling exponents $1/\psi=1.00(1), \zeta =
0.53(1)$ in the small-$D\xi(t)$ scaling regime and $1/\psi =
1.80(2), \zeta = 0.68(1)$ in the large-$D\xi(t)$ scaling regime. The
results indicate that the former belongs to the universality class
of the random depositions, while the later belongs to the
universality class of the QEW equation. The two universality classes
are separated by the so-called Larkin length $L_p$ at which the
effects of the quenched disorder and domain-wall elasticity are of
the same order \cite{tan04,fed05, nog08,kag11}. According to
Eq.~(\ref{equ210}), $L_p \approx 0.5L_c$ is estimated for the creep
dynamics. Now let us recall the growth process of the creep
correlation length. At the beginning, the elasticity is dominant.
The kinetic roughening of the domain interface is then dominated by
thermal fluctuations with $\zeta_T = 1/2$ \cite{due05,kol05}, and
the energy barrier is linear with $D\xi(t)$. After $D\xi(t)$ reaches
$L_p$, the quenched disorder overcomes the elasticity. Then the
domain-wall motion can be described by the QEW equation with the
nontrivial exponents $\psi=1/2$ and $\zeta_{eq} = 2/3$
\cite{kol05a,igu09}.

In Table.~\ref{t2}, the effects of $\Delta$, $H_0$ and $f$ are uncovered in the large-$D\xi(t)$ regime
with the fixed sets of model parameters
($T=0.33, H_0 =0.01, f=10^{-4}$Hz), ($T=0.33, \Delta =1.5, f=10^{-4}$Hz) and
($T=0.33, \Delta =1.5, H_0=0.1$), respectively. For a moderate disorder, i.e.,
$0.5 \leq \Delta \leq 1.5$, a robust value $\psi = 0.54(1)$ is determined,
close to $\psi = 1/2$ of the QEW equation. According to Eq.~(\ref{equ220}), the hopping time
$t \sim \exp\left(\left[D\xi/L_p\right]^{\psi}/T\right)$ is derived with $L_p \sim 1/ \Delta$ \cite{mon08,kag11},
consistent with the ones in Refs.\cite{kol05a, cor11}. The factor $\Delta^{\psi}/T$ shows that the hopping process is determined by the competition between the
quenched disorder and thermal noise \cite{kol05}. Since $1/\psi$ increases monotonically with $H_0$ and the curve $D\xi(t)$ drops
at the tail, further studies are needed to derive the exact functional form on $H_0$ and $f$.

For the kinetic roughening of the domain interface, a robust value
$\zeta = 0.68(1)$ is determined in the large-$D\xi(t)$ scaling
regime at different $T$, $\Delta$, $H_0$ and $f$. The scaling
relation $\zeta = \zeta_{loc} < 1$ indicates that the domain
interfaces belongs to the Family-Vicsek universality class
\cite{ram00}. When $D\xi(t)$ exceeds a certain threshold, however,
$\zeta$ differs from $\zeta_{loc}$ by more than $15$ percent not
only at a higher temperature $T=0.66$, but also at weaker quenched
disorders $\Delta=0.1, 0.5$, stronger driving fields $H_0=0.2, 0.5$
and higher frequencies $f=10^{-1}$ Hz, $1$ Hz. A similar phenomenon
has also been observed in Ref.\cite{kol09} where a crossover of the
roughness exponent from $\zeta_{loc}=2/3$ to $\zeta=1.25$ is
obtained for different driving fields. It suggests that in this case
the domain interface is no longer single-valued and one-dimensional.
As a consequence, the domain interface belongs to a new universality
class with intrinsic anomalous scaling and spatial multiscaling
\cite{zho09,ram00}.

\section{Conclusion}
With Monte Carlo simulations, we have explored the nonsteady
dynamics properties of a domain wall for the creep state under an
alternating driving field. Since the phenomenological QEW equation
contains little microscopic information, lattice models based on
microscopic structures and interactions are considered. Taking the
$2$D DRFIM model as an example, two distinct growth stages of the
domain interface are identified with both the creep correlation
length $D\xi(t)$ and the pure roughness function $D\omega^2(t)$. The
small-$D\xi(t)$ regime corresponds to the universality class of the
random depositions with the scaling exponents $\psi = 1$ and
$\zeta_{T} = 1/2$, while the large-$D\xi(t)$ one belongs to the
universality class of the QEW equation with $\psi = 1/2$ and
$\zeta_{eq} = 2/3$.

However, due to the dynamic effect of overhangs, the roughness exponent $\zeta$ may significantly deviate from
$2/3$ of the QEW equation at either higher temperatures, weaker quenched disorders, stronger driving fields
or higher frequencies, but comparable with experiments \cite{lem98, lee09}.
The result $\zeta > \zeta_{loc} = 2/3$ indicates that the domain interface belongs
to a new universality class with intrinsic anomalous scaling and spatial multiscaling.

In addition, as the temperature increases, the creep exponent $\beta$ measured from the
stationary magnetic hysteresis loops changes from $0.23(1)$ to
$0.55(2)$, compatible with the experimental measurements \cite{bra05,kle07,kle06a}. The temperature-independent
scaling relation $\beta = k \zeta / \psi$ is then observed with the coefficient $k = 0.45(3)$.
With this scaling relation, one can predict the creep exponent $\beta$ by only measuring $\psi$ and $\zeta$
from the nonsteady dynamics.

{\bf Acknowledgements:} This work was supported in part by the National Natural
Science Foundation of China (under Grant Nos. 11205043, 11375149 and 11304072), the Zhejiang Provincial
Natural Science Foundation (under Grants No. LQ12A05002), and the funds from Hangzhou City
for supporting Hangzhou-City Quantum Information and Quantum Optics Innovation Research Team.

\bibliography{zheng,domain}

\begin{thebibliography}{63}%
\makeatletter
\providecommand \@ifxundefined [1]{%
 \@ifx{#1\undefined}
}%
\providecommand \@ifnum [1]{%
 \ifnum #1\expandafter \@firstoftwo
 \else \expandafter \@secondoftwo
 \fi
}%
\providecommand \@ifx [1]{%
 \ifx #1\expandafter \@firstoftwo
 \else \expandafter \@secondoftwo
 \fi
}%
\providecommand \natexlab [1]{#1}%
\providecommand \enquote  [1]{``#1''}%
\providecommand \bibnamefont  [1]{#1}%
\providecommand \bibfnamefont [1]{#1}%
\providecommand \citenamefont [1]{#1}%
\providecommand \href@noop [0]{\@secondoftwo}%
\providecommand \href [0]{\begingroup \@sanitize@url \@href}%
\providecommand \@href[1]{\@@startlink{#1}\@@href}%
\providecommand \@@href[1]{\endgroup#1\@@endlink}%
\providecommand \@sanitize@url [0]{\catcode `\\12\catcode `\$12\catcode
  `\&12\catcode `\#12\catcode `\^12\catcode `\_12\catcode `\%12\relax}%
\providecommand \@@startlink[1]{}%
\providecommand \@@endlink[0]{}%
\providecommand \url  [0]{\begingroup\@sanitize@url \@url }%
\providecommand \@url [1]{\endgroup\@href {#1}{\urlprefix }}%
\providecommand \urlprefix  [0]{URL }%
\providecommand \Eprint [0]{\href }%
\providecommand \doibase [0]{http://dx.doi.org/}%
\providecommand \selectlanguage [0]{\@gobble}%
\providecommand \bibinfo  [0]{\@secondoftwo}%
\providecommand \bibfield  [0]{\@secondoftwo}%
\providecommand \translation [1]{[#1]}%
\providecommand \BibitemOpen [0]{}%
\providecommand \bibitemStop [0]{}%
\providecommand \bibitemNoStop [0]{.\EOS\space}%
\providecommand \EOS [0]{\spacefactor3000\relax}%
\providecommand \BibitemShut  [1]{\csname bibitem#1\endcsname}%
\let\auto@bib@innerbib\@empty
\bibitem [{\citenamefont {Tatara}\ and\ \citenamefont {Kohno}(2004)}]{tat04}%
  \BibitemOpen
  \bibfield  {author} {\bibinfo {author} {\bibfnamefont {G.}~\bibnamefont
  {Tatara}}\ and\ \bibinfo {author} {\bibfnamefont {H.}~\bibnamefont {Kohno}},\
  }\href@noop {} {\bibfield  {journal} {\bibinfo  {journal} {Phys. Rev. Lett.}\
  }\textbf {\bibinfo {volume} {92}},\ \bibinfo {pages} {086601} (\bibinfo
  {year} {2004})}\BibitemShut {NoStop}%
\bibitem [{\citenamefont {Kleemann}(2007)}]{kle07a}%
  \BibitemOpen
  \bibfield  {author} {\bibinfo {author} {\bibfnamefont {W.}~\bibnamefont
  {Kleemann}},\ }\href@noop {} {\bibfield  {journal} {\bibinfo  {journal}
  {Annu. Rev. Mater. Res.}\ }\textbf {\bibinfo {volume} {37}},\ \bibinfo
  {pages} {415} (\bibinfo {year} {2007})}\BibitemShut {NoStop}%
\bibitem [{\citenamefont {Martinez}\ \emph {et~al.}(2007)\citenamefont
  {Martinez}, \citenamefont {Diaz}, \citenamefont {Torres}, \citenamefont
  {Tristan},\ and\ \citenamefont {Alejos}}]{mar07}%
  \BibitemOpen
  \bibfield  {author} {\bibinfo {author} {\bibfnamefont {E.}~\bibnamefont
  {Martinez}}, \bibinfo {author} {\bibfnamefont {L.~L.}\ \bibnamefont {Diaz}},
  \bibinfo {author} {\bibfnamefont {L.}~\bibnamefont {Torres}}, \bibinfo
  {author} {\bibfnamefont {C.}~\bibnamefont {Tristan}}, \ and\ \bibinfo
  {author} {\bibfnamefont {O.}~\bibnamefont {Alejos}},\ }\href@noop {}
  {\bibfield  {journal} {\bibinfo  {journal} {Phys. Rev. B}\ }\textbf {\bibinfo
  {volume} {75}},\ \bibinfo {pages} {174409} (\bibinfo {year}
  {2007})}\BibitemShut {NoStop}%
\bibitem [{\citenamefont {Metaxas}\ \emph {et~al.}(2007)\citenamefont
  {Metaxas}, \citenamefont {Jamet}, \citenamefont {Mougin}, \citenamefont
  {Cormier}, \citenamefont {Ferr\'e}, \citenamefont {Baltz}, \citenamefont
  {Rodmacq}, \citenamefont {Dieny},\ and\ \citenamefont {Stamps}}]{met07}%
  \BibitemOpen
  \bibfield  {author} {\bibinfo {author} {\bibfnamefont {P.~J.}\ \bibnamefont
  {Metaxas}}, \bibinfo {author} {\bibfnamefont {J.~P.}\ \bibnamefont {Jamet}},
  \bibinfo {author} {\bibfnamefont {A.}~\bibnamefont {Mougin}}, \bibinfo
  {author} {\bibfnamefont {M.}~\bibnamefont {Cormier}}, \bibinfo {author}
  {\bibfnamefont {J.}~\bibnamefont {Ferr\'e}}, \bibinfo {author} {\bibfnamefont
  {V.}~\bibnamefont {Baltz}}, \bibinfo {author} {\bibfnamefont
  {B.}~\bibnamefont {Rodmacq}}, \bibinfo {author} {\bibfnamefont
  {B.}~\bibnamefont {Dieny}}, \ and\ \bibinfo {author} {\bibfnamefont {R.~L.}\
  \bibnamefont {Stamps}},\ }\href@noop {} {\bibfield  {journal} {\bibinfo
  {journal} {Phys. Rev. Lett.}\ }\textbf {\bibinfo {volume} {99}},\ \bibinfo
  {pages} {217208} (\bibinfo {year} {2007})}\BibitemShut {NoStop}%
\bibitem [{\citenamefont {Kim}\ \emph {et~al.}(2009)\citenamefont {Kim},
  \citenamefont {Lee}, \citenamefont {Ahn}, \citenamefont {Lee}, \citenamefont
  {Lee}, \citenamefont {Cho}, \citenamefont {Seo}, \citenamefont {Shin},
  \citenamefont {Choe},\ and\ \citenamefont {Lee}}]{kim09}%
  \BibitemOpen
  \bibfield  {author} {\bibinfo {author} {\bibfnamefont {K.~J.}\ \bibnamefont
  {Kim}}, \bibinfo {author} {\bibfnamefont {J.~C.}\ \bibnamefont {Lee}},
  \bibinfo {author} {\bibfnamefont {S.~M.}\ \bibnamefont {Ahn}}, \bibinfo
  {author} {\bibfnamefont {K.~S.}\ \bibnamefont {Lee}}, \bibinfo {author}
  {\bibfnamefont {C.~W.}\ \bibnamefont {Lee}}, \bibinfo {author} {\bibfnamefont
  {Y.~J.}\ \bibnamefont {Cho}}, \bibinfo {author} {\bibfnamefont
  {S.}~\bibnamefont {Seo}}, \bibinfo {author} {\bibfnamefont {K.~H.}\
  \bibnamefont {Shin}}, \bibinfo {author} {\bibfnamefont {S.~B.}\ \bibnamefont
  {Choe}}, \ and\ \bibinfo {author} {\bibfnamefont {H.~W.}\ \bibnamefont
  {Lee}},\ }\href@noop {} {\bibfield  {journal} {\bibinfo  {journal} {Nature}\
  }\textbf {\bibinfo {volume} {458}},\ \bibinfo {pages} {740} (\bibinfo {year}
  {2009})}\BibitemShut {NoStop}%
\bibitem [{\citenamefont {Sch\"{u}tze}\ and\ \citenamefont
  {Nattermann}(2011)}]{sch11}%
  \BibitemOpen
  \bibfield  {author} {\bibinfo {author} {\bibfnamefont {F.}~\bibnamefont
  {Sch\"{u}tze}}\ and\ \bibinfo {author} {\bibfnamefont {T.}~\bibnamefont
  {Nattermann}},\ }\href@noop {} {\bibfield  {journal} {\bibinfo  {journal}
  {Phys. Rev. B}\ }\textbf {\bibinfo {volume} {83}},\ \bibinfo {pages} {024412}
  (\bibinfo {year} {2011})}\BibitemShut {NoStop}%
\bibitem [{\citenamefont {Miron}\ \emph {et~al.}(2011)\citenamefont {Miron},
  \citenamefont {Moore}, \citenamefont {Szambolics}, \citenamefont {Prejbeanu},
  \citenamefont {Auffret}, \citenamefont {Rodmacq}, \citenamefont {Pizzini},
  \citenamefont {Vogel}, \citenamefont {Bonfim}, \citenamefont {Schuhl},\ and\
  \citenamefont {Gaudin}}]{mir11}%
  \BibitemOpen
  \bibfield  {author} {\bibinfo {author} {\bibfnamefont {I.~M.}\ \bibnamefont
  {Miron}}, \bibinfo {author} {\bibfnamefont {T.}~\bibnamefont {Moore}},
  \bibinfo {author} {\bibfnamefont {H.}~\bibnamefont {Szambolics}}, \bibinfo
  {author} {\bibfnamefont {L.~D.~B.}\ \bibnamefont {Prejbeanu}}, \bibinfo
  {author} {\bibfnamefont {S.}~\bibnamefont {Auffret}}, \bibinfo {author}
  {\bibfnamefont {B.}~\bibnamefont {Rodmacq}}, \bibinfo {author} {\bibfnamefont
  {S.}~\bibnamefont {Pizzini}}, \bibinfo {author} {\bibfnamefont
  {J.}~\bibnamefont {Vogel}}, \bibinfo {author} {\bibfnamefont
  {M.}~\bibnamefont {Bonfim}}, \bibinfo {author} {\bibfnamefont
  {A.}~\bibnamefont {Schuhl}}, \ and\ \bibinfo {author} {\bibfnamefont
  {G.}~\bibnamefont {Gaudin}},\ }\href@noop {} {\bibfield  {journal} {\bibinfo
  {journal} {Nature Mater.}\ }\textbf {\bibinfo {volume} {10}},\ \bibinfo
  {pages} {419} (\bibinfo {year} {2011})}\BibitemShut {NoStop}%
\bibitem [{\citenamefont {Lee}\ \emph {et~al.}(2011)\citenamefont {Lee},
  \citenamefont {Kim}, \citenamefont {Ryu}, \citenamefont {Moon}, \citenamefont
  {Yun}, \citenamefont {Gim}, \citenamefont {Lee}, \citenamefont {Shin},
  \citenamefont {Lee},\ and\ \citenamefont {Choe}}]{lee11}%
  \BibitemOpen
  \bibfield  {author} {\bibinfo {author} {\bibfnamefont {J.~C.}\ \bibnamefont
  {Lee}}, \bibinfo {author} {\bibfnamefont {K.~J.}\ \bibnamefont {Kim}},
  \bibinfo {author} {\bibfnamefont {J.}~\bibnamefont {Ryu}}, \bibinfo {author}
  {\bibfnamefont {K.~W.}\ \bibnamefont {Moon}}, \bibinfo {author}
  {\bibfnamefont {S.~J.}\ \bibnamefont {Yun}}, \bibinfo {author} {\bibfnamefont
  {G.~H.}\ \bibnamefont {Gim}}, \bibinfo {author} {\bibfnamefont {K.~S.}\
  \bibnamefont {Lee}}, \bibinfo {author} {\bibfnamefont {K.~H.}\ \bibnamefont
  {Shin}}, \bibinfo {author} {\bibfnamefont {H.~W.}\ \bibnamefont {Lee}}, \
  and\ \bibinfo {author} {\bibfnamefont {S.~B.}\ \bibnamefont {Choe}},\
  }\href@noop {} {\bibfield  {journal} {\bibinfo  {journal} {Phys. Rev. Lett.}\
  }\textbf {\bibinfo {volume} {107}},\ \bibinfo {pages} {067201} (\bibinfo
  {year} {2011})}\BibitemShut {NoStop}%
\bibitem [{\citenamefont {Dolz}\ \emph {et~al.}(2010)\citenamefont {Dolz},
  \citenamefont {Kolton},\ and\ \citenamefont {Pastoriza}}]{dol10}%
  \BibitemOpen
  \bibfield  {author} {\bibinfo {author} {\bibfnamefont {M.~I.}\ \bibnamefont
  {Dolz}}, \bibinfo {author} {\bibfnamefont {A.~B.}\ \bibnamefont {Kolton}}, \
  and\ \bibinfo {author} {\bibfnamefont {H.}~\bibnamefont {Pastoriza}},\
  }\href@noop {} {\bibfield  {journal} {\bibinfo  {journal} {Phys. Rev. B}\
  }\textbf {\bibinfo {volume} {81}},\ \bibinfo {pages} {092502} (\bibinfo
  {year} {2010})}\BibitemShut {NoStop}%
\bibitem [{\citenamefont {Cao}\ \emph {et~al.}(2012)\citenamefont {Cao},
  \citenamefont {Luo},\ and\ \citenamefont {Hu}}]{cao12}%
  \BibitemOpen
  \bibfield  {author} {\bibinfo {author} {\bibfnamefont {W.~P.}\ \bibnamefont
  {Cao}}, \bibinfo {author} {\bibfnamefont {M.~B.}\ \bibnamefont {Luo}}, \ and\
  \bibinfo {author} {\bibfnamefont {X.}~\bibnamefont {Hu}},\ }\href@noop {}
  {\bibfield  {journal} {\bibinfo  {journal} {New J. Phys.}\ }\textbf {\bibinfo
  {volume} {14}},\ \bibinfo {pages} {013006} (\bibinfo {year}
  {2012})}\BibitemShut {NoStop}%
\bibitem [{\citenamefont {Je\.{z}ewski}\ \emph {et~al.}(2008)\citenamefont
  {Je\.{z}ewski}, \citenamefont {Kuczy\'{n}ski},\ and\ \citenamefont
  {Hoffmann}}]{jez08}%
  \BibitemOpen
  \bibfield  {author} {\bibinfo {author} {\bibfnamefont {W.}~\bibnamefont
  {Je\.{z}ewski}}, \bibinfo {author} {\bibfnamefont {W.}~\bibnamefont
  {Kuczy\'{n}ski}}, \ and\ \bibinfo {author} {\bibfnamefont {J.}~\bibnamefont
  {Hoffmann}},\ }\href@noop {} {\bibfield  {journal} {\bibinfo  {journal}
  {Phys. Rev. B}\ }\textbf {\bibinfo {volume} {77}},\ \bibinfo {pages} {094101}
  (\bibinfo {year} {2008})}\BibitemShut {NoStop}%
\bibitem [{\citenamefont {Kagawa}\ \emph {et~al.}(2011)\citenamefont {Kagawa},
  \citenamefont {Onose}, \citenamefont {Kaneko},\ and\ \citenamefont
  {Tokura}}]{kag11}%
  \BibitemOpen
  \bibfield  {author} {\bibinfo {author} {\bibfnamefont {F.}~\bibnamefont
  {Kagawa}}, \bibinfo {author} {\bibfnamefont {Y.}~\bibnamefont {Onose}},
  \bibinfo {author} {\bibfnamefont {Y.}~\bibnamefont {Kaneko}}, \ and\ \bibinfo
  {author} {\bibfnamefont {Y.}~\bibnamefont {Tokura}},\ }\href@noop {}
  {\bibfield  {journal} {\bibinfo  {journal} {Phys. Rev. B}\ }\textbf {\bibinfo
  {volume} {83}},\ \bibinfo {pages} {054413} (\bibinfo {year}
  {2011})}\BibitemShut {NoStop}%
\bibitem [{\citenamefont {Steinke}\ \emph {et~al.}(2012)\citenamefont
  {Steinke}, \citenamefont {Moore}, \citenamefont {Mansell}, \citenamefont
  {Bland},\ and\ \citenamefont {Barnes}}]{ste12}%
  \BibitemOpen
  \bibfield  {author} {\bibinfo {author} {\bibfnamefont {N.~J.}\ \bibnamefont
  {Steinke}}, \bibinfo {author} {\bibfnamefont {T.~A.}\ \bibnamefont {Moore}},
  \bibinfo {author} {\bibfnamefont {R.}~\bibnamefont {Mansell}}, \bibinfo
  {author} {\bibfnamefont {J.~A.~C.}\ \bibnamefont {Bland}}, \ and\ \bibinfo
  {author} {\bibfnamefont {C.~H.~W.}\ \bibnamefont {Barnes}},\ }\href@noop {}
  {\bibfield  {journal} {\bibinfo  {journal} {Phys. Rev. B}\ }\textbf {\bibinfo
  {volume} {86}},\ \bibinfo {pages} {184434} (\bibinfo {year}
  {2012})}\BibitemShut {NoStop}%
\bibitem [{\citenamefont {Zhernenkov}\ \emph {et~al.}(2013)\citenamefont
  {Zhernenkov}, \citenamefont {Gorkov}, \citenamefont {Toperverg},\ and\
  \citenamefont {Zabel}}]{zhe13}%
  \BibitemOpen
  \bibfield  {author} {\bibinfo {author} {\bibfnamefont {K.}~\bibnamefont
  {Zhernenkov}}, \bibinfo {author} {\bibfnamefont {D.}~\bibnamefont {Gorkov}},
  \bibinfo {author} {\bibfnamefont {B.~P.}\ \bibnamefont {Toperverg}}, \ and\
  \bibinfo {author} {\bibfnamefont {H.}~\bibnamefont {Zabel}},\ }\href@noop {}
  {\bibfield  {journal} {\bibinfo  {journal} {Phys. Rev. B}\ }\textbf {\bibinfo
  {volume} {88}},\ \bibinfo {pages} {020401(R)} (\bibinfo {year}
  {2013})}\BibitemShut {NoStop}%
\bibitem [{\citenamefont {Laurson}\ and\ \citenamefont {Alava}(2012)}]{lau12}%
  \BibitemOpen
  \bibfield  {author} {\bibinfo {author} {\bibfnamefont {L.}~\bibnamefont
  {Laurson}}\ and\ \bibinfo {author} {\bibfnamefont {M.~J.}\ \bibnamefont
  {Alava}},\ }\href@noop {} {\bibfield  {journal} {\bibinfo  {journal} {Phys.
  Rev. Lett.}\ }\textbf {\bibinfo {volume} {109}},\ \bibinfo {pages} {155504}
  (\bibinfo {year} {2012})}\BibitemShut {NoStop}%
\bibitem [{\citenamefont {Yamanouchi}\ \emph {et~al.}(2007)\citenamefont
  {Yamanouchi}, \citenamefont {Ieda}, \citenamefont {Matsukura}, \citenamefont
  {Barnes}, \citenamefont {Maekawa},\ and\ \citenamefont {Ohno}}]{yam07}%
  \BibitemOpen
  \bibfield  {author} {\bibinfo {author} {\bibfnamefont {M.}~\bibnamefont
  {Yamanouchi}}, \bibinfo {author} {\bibfnamefont {J.}~\bibnamefont {Ieda}},
  \bibinfo {author} {\bibfnamefont {F.}~\bibnamefont {Matsukura}}, \bibinfo
  {author} {\bibfnamefont {S.~E.}\ \bibnamefont {Barnes}}, \bibinfo {author}
  {\bibfnamefont {S.}~\bibnamefont {Maekawa}}, \ and\ \bibinfo {author}
  {\bibfnamefont {H.}~\bibnamefont {Ohno}},\ }\href@noop {} {\bibfield
  {journal} {\bibinfo  {journal} {Science}\ }\textbf {\bibinfo {volume}
  {317}},\ \bibinfo {pages} {1726} (\bibinfo {year} {2007})}\BibitemShut
  {NoStop}%
\bibitem [{\citenamefont {Hayashi}\ \emph {et~al.}(2008)\citenamefont
  {Hayashi}, \citenamefont {Thomas}, \citenamefont {Moriya}, \citenamefont
  {Rettner},\ and\ \citenamefont {Parkin}}]{hay08}%
  \BibitemOpen
  \bibfield  {author} {\bibinfo {author} {\bibfnamefont {M.}~\bibnamefont
  {Hayashi}}, \bibinfo {author} {\bibfnamefont {L.}~\bibnamefont {Thomas}},
  \bibinfo {author} {\bibfnamefont {R.}~\bibnamefont {Moriya}}, \bibinfo
  {author} {\bibfnamefont {C.}~\bibnamefont {Rettner}}, \ and\ \bibinfo
  {author} {\bibfnamefont {S.~S.~P.}\ \bibnamefont {Parkin}},\ }\href@noop {}
  {\bibfield  {journal} {\bibinfo  {journal} {Science}\ }\textbf {\bibinfo
  {volume} {320}},\ \bibinfo {pages} {209} (\bibinfo {year}
  {2008})}\BibitemShut {NoStop}%
\bibitem [{\citenamefont {Parkin}\ \emph {et~al.}(2008)\citenamefont {Parkin},
  \citenamefont {Hayashi},\ and\ \citenamefont {Thomas}}]{par08}%
  \BibitemOpen
  \bibfield  {author} {\bibinfo {author} {\bibfnamefont {S.~S.~P.}\
  \bibnamefont {Parkin}}, \bibinfo {author} {\bibfnamefont {M.}~\bibnamefont
  {Hayashi}}, \ and\ \bibinfo {author} {\bibfnamefont {L.}~\bibnamefont
  {Thomas}},\ }\href@noop {} {\bibfield  {journal} {\bibinfo  {journal}
  {Science}\ }\textbf {\bibinfo {volume} {320}},\ \bibinfo {pages} {190}
  (\bibinfo {year} {2008})}\BibitemShut {NoStop}%
\bibitem [{\citenamefont {Venimadhav}\ \emph {et~al.}(2013)\citenamefont
  {Venimadhav}, \citenamefont {Chandrasekar},\ and\ \citenamefont
  {Murthy}}]{ven13}%
  \BibitemOpen
  \bibfield  {author} {\bibinfo {author} {\bibfnamefont {A.}~\bibnamefont
  {Venimadhav}}, \bibinfo {author} {\bibfnamefont {D.}~\bibnamefont
  {Chandrasekar}}, \ and\ \bibinfo {author} {\bibfnamefont {J.~K.}\
  \bibnamefont {Murthy}},\ }\href@noop {} {\bibfield  {journal} {\bibinfo
  {journal} {Appl. Nanosci.}\ }\textbf {\bibinfo {volume} {3}},\ \bibinfo
  {pages} {25} (\bibinfo {year} {2013})}\BibitemShut {NoStop}%
\bibitem [{\citenamefont {Braun}\ \emph {et~al.}(2005)\citenamefont {Braun},
  \citenamefont {Kleemann}, \citenamefont {Dec},\ and\ \citenamefont
  {Thomas}}]{bra05}%
  \BibitemOpen
  \bibfield  {author} {\bibinfo {author} {\bibfnamefont {T.}~\bibnamefont
  {Braun}}, \bibinfo {author} {\bibfnamefont {W.}~\bibnamefont {Kleemann}},
  \bibinfo {author} {\bibfnamefont {J.}~\bibnamefont {Dec}}, \ and\ \bibinfo
  {author} {\bibfnamefont {P.~A.}\ \bibnamefont {Thomas}},\ }\href@noop {}
  {\bibfield  {journal} {\bibinfo  {journal} {Phys. Rev. Lett.}\ }\textbf
  {\bibinfo {volume} {94}},\ \bibinfo {pages} {117601} (\bibinfo {year}
  {2005})}\BibitemShut {NoStop}%
\bibitem [{\citenamefont {Kleemann}\ \emph {et~al.}(2007)\citenamefont
  {Kleemann}, \citenamefont {Rhensius}, \citenamefont {Petracic}, \citenamefont
  {Ferr\'e}, \citenamefont {Jamet},\ and\ \citenamefont {Bernas}}]{kle07}%
  \BibitemOpen
  \bibfield  {author} {\bibinfo {author} {\bibfnamefont {W.}~\bibnamefont
  {Kleemann}}, \bibinfo {author} {\bibfnamefont {J.}~\bibnamefont {Rhensius}},
  \bibinfo {author} {\bibfnamefont {O.}~\bibnamefont {Petracic}}, \bibinfo
  {author} {\bibfnamefont {J.}~\bibnamefont {Ferr\'e}}, \bibinfo {author}
  {\bibfnamefont {J.~P.}\ \bibnamefont {Jamet}}, \ and\ \bibinfo {author}
  {\bibfnamefont {H.}~\bibnamefont {Bernas}},\ }\href@noop {} {\bibfield
  {journal} {\bibinfo  {journal} {Phys. Rev. Lett.}\ }\textbf {\bibinfo
  {volume} {99}},\ \bibinfo {pages} {097203} (\bibinfo {year}
  {2007})}\BibitemShut {NoStop}%
\bibitem [{\citenamefont {Kleemann}\ \emph {et~al.}(2006)\citenamefont
  {Kleemann}, \citenamefont {Dec}, \citenamefont {Prosandeev}, \citenamefont
  {Braun},\ and\ \citenamefont {Thomas}}]{kle06a}%
  \BibitemOpen
  \bibfield  {author} {\bibinfo {author} {\bibfnamefont {W.}~\bibnamefont
  {Kleemann}}, \bibinfo {author} {\bibfnamefont {J.}~\bibnamefont {Dec}},
  \bibinfo {author} {\bibfnamefont {S.~A.}\ \bibnamefont {Prosandeev}},
  \bibinfo {author} {\bibfnamefont {T.}~\bibnamefont {Braun}}, \ and\ \bibinfo
  {author} {\bibfnamefont {P.~A.}\ \bibnamefont {Thomas}},\ }\href@noop {}
  {\bibfield  {journal} {\bibinfo  {journal} {Ferroelectrics}\ }\textbf
  {\bibinfo {volume} {334}},\ \bibinfo {pages} {3} (\bibinfo {year}
  {2006})}\BibitemShut {NoStop}%
\bibitem [{\citenamefont {Yang}\ \emph {et~al.}(2010)\citenamefont {Yang},
  \citenamefont {Jo}, \citenamefont {Kim}, \citenamefont {Yoon}, \citenamefont
  {Song}, \citenamefont {Lee}, \citenamefont {Marton}, \citenamefont {Park},
  \citenamefont {Jo},\ and\ \citenamefont {Noh}}]{yan10}%
  \BibitemOpen
  \bibfield  {author} {\bibinfo {author} {\bibfnamefont {S.~M.}\ \bibnamefont
  {Yang}}, \bibinfo {author} {\bibfnamefont {J.~Y.}\ \bibnamefont {Jo}},
  \bibinfo {author} {\bibfnamefont {T.~H.}\ \bibnamefont {Kim}}, \bibinfo
  {author} {\bibfnamefont {J.~G.}\ \bibnamefont {Yoon}}, \bibinfo {author}
  {\bibfnamefont {T.~K.}\ \bibnamefont {Song}}, \bibinfo {author}
  {\bibfnamefont {H.~N.}\ \bibnamefont {Lee}}, \bibinfo {author} {\bibfnamefont
  {Z.}~\bibnamefont {Marton}}, \bibinfo {author} {\bibfnamefont
  {S.}~\bibnamefont {Park}}, \bibinfo {author} {\bibfnamefont {Y.}~\bibnamefont
  {Jo}}, \ and\ \bibinfo {author} {\bibfnamefont {T.~W.}\ \bibnamefont {Noh}},\
  }\href@noop {} {\bibfield  {journal} {\bibinfo  {journal} {Phys. Rev. B}\
  }\textbf {\bibinfo {volume} {82}},\ \bibinfo {pages} {174125} (\bibinfo
  {year} {2010})}\BibitemShut {NoStop}%
\bibitem [{\citenamefont {Harrison}\ \emph {et~al.}(2004)\citenamefont
  {Harrison}, \citenamefont {Redfern},\ and\ \citenamefont {Salje}}]{har04}%
  \BibitemOpen
  \bibfield  {author} {\bibinfo {author} {\bibfnamefont {R.~J.}\ \bibnamefont
  {Harrison}}, \bibinfo {author} {\bibfnamefont {S.~A.~T.}\ \bibnamefont
  {Redfern}}, \ and\ \bibinfo {author} {\bibfnamefont {E.~K.~H.}\ \bibnamefont
  {Salje}},\ }\href@noop {} {\bibfield  {journal} {\bibinfo  {journal} {Phys.
  Rev. B}\ }\textbf {\bibinfo {volume} {69}},\ \bibinfo {pages} {144101}
  (\bibinfo {year} {2004})}\BibitemShut {NoStop}%
\bibitem [{\citenamefont {Mushenok}\ \emph {et~al.}(2011)\citenamefont
  {Mushenok}, \citenamefont {Koplak},\ and\ \citenamefont {Morgunov}}]{mus11}%
  \BibitemOpen
  \bibfield  {author} {\bibinfo {author} {\bibfnamefont {F.}~\bibnamefont
  {Mushenok}}, \bibinfo {author} {\bibfnamefont {O.}~\bibnamefont {Koplak}}, \
  and\ \bibinfo {author} {\bibfnamefont {R.}~\bibnamefont {Morgunov}},\
  }\href@noop {} {\bibfield  {journal} {\bibinfo  {journal} {Eur. Phys. J. B}\
  }\textbf {\bibinfo {volume} {84}},\ \bibinfo {pages} {219} (\bibinfo {year}
  {2011})}\BibitemShut {NoStop}%
\bibitem [{\citenamefont {Nattermann}\ \emph {et~al.}(2001)\citenamefont
  {Nattermann}, \citenamefont {Pokrovsky},\ and\ \citenamefont
  {Vinokur}}]{nat01}%
  \BibitemOpen
  \bibfield  {author} {\bibinfo {author} {\bibfnamefont {T.}~\bibnamefont
  {Nattermann}}, \bibinfo {author} {\bibfnamefont {V.}~\bibnamefont
  {Pokrovsky}}, \ and\ \bibinfo {author} {\bibfnamefont {V.~M.}\ \bibnamefont
  {Vinokur}},\ }\href@noop {} {\bibfield  {journal} {\bibinfo  {journal} {Phys.
  Rev. Lett.}\ }\textbf {\bibinfo {volume} {87}},\ \bibinfo {pages} {197005}
  (\bibinfo {year} {2001})}\BibitemShut {NoStop}%
\bibitem [{\citenamefont {Glatz}\ \emph {et~al.}(2003)\citenamefont {Glatz},
  \citenamefont {Nattermann},\ and\ \citenamefont {Pokrovsky}}]{gla03}%
  \BibitemOpen
  \bibfield  {author} {\bibinfo {author} {\bibfnamefont {A.}~\bibnamefont
  {Glatz}}, \bibinfo {author} {\bibfnamefont {T.}~\bibnamefont {Nattermann}}, \
  and\ \bibinfo {author} {\bibfnamefont {V.}~\bibnamefont {Pokrovsky}},\
  }\href@noop {} {\bibfield  {journal} {\bibinfo  {journal} {Phys. Rev. Lett.}\
  }\textbf {\bibinfo {volume} {90}},\ \bibinfo {pages} {047201} (\bibinfo
  {year} {2003})}\BibitemShut {NoStop}%
\bibitem [{\citenamefont {Petracic}\ \emph {et~al.}(2004)\citenamefont
  {Petracic}, \citenamefont {Glatz},\ and\ \citenamefont {Kleemann}}]{pet04}%
  \BibitemOpen
  \bibfield  {author} {\bibinfo {author} {\bibfnamefont {O.}~\bibnamefont
  {Petracic}}, \bibinfo {author} {\bibfnamefont {A.}~\bibnamefont {Glatz}}, \
  and\ \bibinfo {author} {\bibfnamefont {W.}~\bibnamefont {Kleemann}},\
  }\href@noop {} {\bibfield  {journal} {\bibinfo  {journal} {Phys. Rev. B}\
  }\textbf {\bibinfo {volume} {70}},\ \bibinfo {pages} {214432} (\bibinfo
  {year} {2004})}\BibitemShut {NoStop}%
\bibitem [{\citenamefont {Kolton}\ \emph {et~al.}(2006)\citenamefont {Kolton},
  \citenamefont {Rosso}, \citenamefont {Albano},\ and\ \citenamefont
  {Giamarchi}}]{kol06}%
  \BibitemOpen
  \bibfield  {author} {\bibinfo {author} {\bibfnamefont {A.~B.}\ \bibnamefont
  {Kolton}}, \bibinfo {author} {\bibfnamefont {A.}~\bibnamefont {Rosso}},
  \bibinfo {author} {\bibfnamefont {E.~V.}\ \bibnamefont {Albano}}, \ and\
  \bibinfo {author} {\bibfnamefont {T.}~\bibnamefont {Giamarchi}},\ }\href@noop
  {} {\bibfield  {journal} {\bibinfo  {journal} {Phys. Rev. B}\ }\textbf
  {\bibinfo {volume} {74}},\ \bibinfo {pages} {140201(R)} (\bibinfo {year}
  {2006})}\BibitemShut {NoStop}%
\bibitem [{\citenamefont {Duemmer}\ and\ \citenamefont {Krauth}(2005)}]{due05}%
  \BibitemOpen
  \bibfield  {author} {\bibinfo {author} {\bibfnamefont {O.}~\bibnamefont
  {Duemmer}}\ and\ \bibinfo {author} {\bibfnamefont {W.}~\bibnamefont
  {Krauth}},\ }\href@noop {} {\bibfield  {journal} {\bibinfo  {journal} {Phys.
  Rev. E}\ }\textbf {\bibinfo {volume} {71}},\ \bibinfo {pages} {061601}
  (\bibinfo {year} {2005})}\BibitemShut {NoStop}%
\bibitem [{\citenamefont {Ferrero}\ \emph {et~al.}(2013)\citenamefont
  {Ferrero}, \citenamefont {Bustingorry},\ and\ \citenamefont
  {Kolton}}]{fer13}%
  \BibitemOpen
  \bibfield  {author} {\bibinfo {author} {\bibfnamefont {E.~E.}\ \bibnamefont
  {Ferrero}}, \bibinfo {author} {\bibfnamefont {S.}~\bibnamefont
  {Bustingorry}}, \ and\ \bibinfo {author} {\bibfnamefont {A.~B.}\ \bibnamefont
  {Kolton}},\ }\href@noop {} {\bibfield  {journal} {\bibinfo  {journal} {Phys.
  Rev. E}\ }\textbf {\bibinfo {volume} {87}},\ \bibinfo {pages} {032122}
  (\bibinfo {year} {2013})}\BibitemShut {NoStop}%
\bibitem [{\citenamefont {Sch\"utze}(2010)}]{sch10}%
  \BibitemOpen
  \bibfield  {author} {\bibinfo {author} {\bibfnamefont {F.}~\bibnamefont
  {Sch\"utze}},\ }\href@noop {} {\bibfield  {journal} {\bibinfo  {journal}
  {Phys. Rev. E}\ }\textbf {\bibinfo {volume} {81}},\ \bibinfo {pages} {051128}
  (\bibinfo {year} {2010})}\BibitemShut {NoStop}%
\bibitem [{\citenamefont {Kolton}\ \emph
  {et~al.}(2005{\natexlab{a}})\citenamefont {Kolton}, \citenamefont {Rosso},\
  and\ \citenamefont {Giamarchi}}]{kol05}%
  \BibitemOpen
  \bibfield  {author} {\bibinfo {author} {\bibfnamefont {A.~B.}\ \bibnamefont
  {Kolton}}, \bibinfo {author} {\bibfnamefont {A.}~\bibnamefont {Rosso}}, \
  and\ \bibinfo {author} {\bibfnamefont {T.}~\bibnamefont {Giamarchi}},\
  }\href@noop {} {\bibfield  {journal} {\bibinfo  {journal} {Phys. Rev. Lett.}\
  }\textbf {\bibinfo {volume} {94}},\ \bibinfo {pages} {047002} (\bibinfo
  {year} {2005}{\natexlab{a}})}\BibitemShut {NoStop}%
\bibitem [{\citenamefont {Kolton}\ \emph {et~al.}(2009)\citenamefont {Kolton},
  \citenamefont {Rosso}, \citenamefont {Giamarchi},\ and\ \citenamefont
  {Krauth}}]{kol09}%
  \BibitemOpen
  \bibfield  {author} {\bibinfo {author} {\bibfnamefont {A.~B.}\ \bibnamefont
  {Kolton}}, \bibinfo {author} {\bibfnamefont {A.}~\bibnamefont {Rosso}},
  \bibinfo {author} {\bibfnamefont {T.}~\bibnamefont {Giamarchi}}, \ and\
  \bibinfo {author} {\bibfnamefont {W.}~\bibnamefont {Krauth}},\ }\href@noop {}
  {\bibfield  {journal} {\bibinfo  {journal} {Phys. Rev. B}\ }\textbf {\bibinfo
  {volume} {79}},\ \bibinfo {pages} {184207} (\bibinfo {year}
  {2009})}\BibitemShut {NoStop}%
\bibitem [{\citenamefont {Chauve}\ \emph {et~al.}(2000)\citenamefont {Chauve},
  \citenamefont {Giamarchi},\ and\ \citenamefont {Doussal}}]{cha00}%
  \BibitemOpen
  \bibfield  {author} {\bibinfo {author} {\bibfnamefont {P.}~\bibnamefont
  {Chauve}}, \bibinfo {author} {\bibfnamefont {T.}~\bibnamefont {Giamarchi}}, \
  and\ \bibinfo {author} {\bibfnamefont {P.~L.}\ \bibnamefont {Doussal}},\
  }\href@noop {} {\bibfield  {journal} {\bibinfo  {journal} {Phys. Rev. B}\
  }\textbf {\bibinfo {volume} {62}},\ \bibinfo {pages} {6241} (\bibinfo {year}
  {2000})}\BibitemShut {NoStop}%
\bibitem [{\citenamefont {Monthus}\ and\ \citenamefont
  {Garel}(2008{\natexlab{a}})}]{mon08}%
  \BibitemOpen
  \bibfield  {author} {\bibinfo {author} {\bibfnamefont {C.}~\bibnamefont
  {Monthus}}\ and\ \bibinfo {author} {\bibfnamefont {T.}~\bibnamefont
  {Garel}},\ }\href@noop {} {\bibfield  {journal} {\bibinfo  {journal} {Phys.
  Rev. E}\ }\textbf {\bibinfo {volume} {78}},\ \bibinfo {pages} {041133}
  (\bibinfo {year} {2008}{\natexlab{a}})}\BibitemShut {NoStop}%
\bibitem [{\citenamefont {Monthus}\ and\ \citenamefont
  {Garel}(2008{\natexlab{b}})}]{mon08b}%
  \BibitemOpen
  \bibfield  {author} {\bibinfo {author} {\bibfnamefont {C.}~\bibnamefont
  {Monthus}}\ and\ \bibinfo {author} {\bibfnamefont {T.}~\bibnamefont
  {Garel}},\ }\href@noop {} {\bibfield  {journal} {\bibinfo  {journal} {J.
  Stat. Mech.}\ }\textbf {\bibinfo {volume} {2008}},\ \bibinfo {pages} {P07002}
  (\bibinfo {year} {2008}{\natexlab{b}})}\BibitemShut {NoStop}%
\bibitem [{\citenamefont {Kolton}\ \emph
  {et~al.}(2005{\natexlab{b}})\citenamefont {Kolton}, \citenamefont {Rosso},\
  and\ \citenamefont {Giamarchi}}]{kol05a}%
  \BibitemOpen
  \bibfield  {author} {\bibinfo {author} {\bibfnamefont {A.~B.}\ \bibnamefont
  {Kolton}}, \bibinfo {author} {\bibfnamefont {A.}~\bibnamefont {Rosso}}, \
  and\ \bibinfo {author} {\bibfnamefont {T.}~\bibnamefont {Giamarchi}},\
  }\href@noop {} {\bibfield  {journal} {\bibinfo  {journal} {Phys. Rev. Lett.}\
  }\textbf {\bibinfo {volume} {95}},\ \bibinfo {pages} {180604} (\bibinfo
  {year} {2005}{\natexlab{b}})}\BibitemShut {NoStop}%
\bibitem [{\citenamefont {Monthus}\ and\ \citenamefont
  {Garel}(2008{\natexlab{c}})}]{mon08a}%
  \BibitemOpen
  \bibfield  {author} {\bibinfo {author} {\bibfnamefont {C.}~\bibnamefont
  {Monthus}}\ and\ \bibinfo {author} {\bibfnamefont {T.}~\bibnamefont
  {Garel}},\ }\href@noop {} {\bibfield  {journal} {\bibinfo  {journal} {J.
  Phys. A: Math. Theor.}\ }\textbf {\bibinfo {volume} {41}},\ \bibinfo {pages}
  {115002} (\bibinfo {year} {2008}{\natexlab{c}})}\BibitemShut {NoStop}%
\bibitem [{\citenamefont {Colaiori}\ \emph {et~al.}(2006)\citenamefont
  {Colaiori}, \citenamefont {Durin},\ and\ \citenamefont {Zapperi}}]{col06}%
  \BibitemOpen
  \bibfield  {author} {\bibinfo {author} {\bibfnamefont {F.}~\bibnamefont
  {Colaiori}}, \bibinfo {author} {\bibfnamefont {G.}~\bibnamefont {Durin}}, \
  and\ \bibinfo {author} {\bibfnamefont {S.}~\bibnamefont {Zapperi}},\
  }\href@noop {} {\bibfield  {journal} {\bibinfo  {journal} {Phys. Rev. Lett.}\
  }\textbf {\bibinfo {volume} {97}},\ \bibinfo {pages} {257203} (\bibinfo
  {year} {2006})}\BibitemShut {NoStop}%
\bibitem [{\citenamefont {Zhou}\ \emph {et~al.}(2009)\citenamefont {Zhou},
  \citenamefont {Zheng},\ and\ \citenamefont {He}}]{zho09}%
  \BibitemOpen
  \bibfield  {author} {\bibinfo {author} {\bibfnamefont {N.~J.}\ \bibnamefont
  {Zhou}}, \bibinfo {author} {\bibfnamefont {B.}~\bibnamefont {Zheng}}, \ and\
  \bibinfo {author} {\bibfnamefont {Y.~Y.}\ \bibnamefont {He}},\ }\href@noop {}
  {\bibfield  {journal} {\bibinfo  {journal} {Phys. Rev. B}\ }\textbf {\bibinfo
  {volume} {80}},\ \bibinfo {pages} {134425} (\bibinfo {year}
  {2009})}\BibitemShut {NoStop}%
\bibitem [{\citenamefont {Zhou}\ \emph {et~al.}(2010)\citenamefont {Zhou},
  \citenamefont {Zheng},\ and\ \citenamefont {Landau}}]{zho10}%
  \BibitemOpen
  \bibfield  {author} {\bibinfo {author} {\bibfnamefont {N.~J.}\ \bibnamefont
  {Zhou}}, \bibinfo {author} {\bibfnamefont {B.}~\bibnamefont {Zheng}}, \ and\
  \bibinfo {author} {\bibfnamefont {D.~P.}\ \bibnamefont {Landau}},\
  }\href@noop {} {\bibfield  {journal} {\bibinfo  {journal} {Europhys. Lett.}\
  }\textbf {\bibinfo {volume} {92}},\ \bibinfo {pages} {36001} (\bibinfo {year}
  {2010})}\BibitemShut {NoStop}%
\bibitem [{\citenamefont {Dong}\ \emph
  {et~al.}(2012{\natexlab{a}})\citenamefont {Dong}, \citenamefont {Zheng},\
  and\ \citenamefont {Zhou}}]{don12a}%
  \BibitemOpen
  \bibfield  {author} {\bibinfo {author} {\bibfnamefont {R.~H.}\ \bibnamefont
  {Dong}}, \bibinfo {author} {\bibfnamefont {B.}~\bibnamefont {Zheng}}, \ and\
  \bibinfo {author} {\bibfnamefont {N.~J.}\ \bibnamefont {Zhou}},\ }\href@noop
  {} {\bibfield  {journal} {\bibinfo  {journal} {Europhys. Lett.}\ }\textbf
  {\bibinfo {volume} {98}},\ \bibinfo {pages} {36002} (\bibinfo {year}
  {2012}{\natexlab{a}})}\BibitemShut {NoStop}%
\bibitem [{\citenamefont {Zhou}\ and\ \citenamefont {Zheng}(2010)}]{zho10a}%
  \BibitemOpen
  \bibfield  {author} {\bibinfo {author} {\bibfnamefont {N.~J.}\ \bibnamefont
  {Zhou}}\ and\ \bibinfo {author} {\bibfnamefont {B.}~\bibnamefont {Zheng}},\
  }\href@noop {} {\bibfield  {journal} {\bibinfo  {journal} {Phys. Rev. E}\
  }\textbf {\bibinfo {volume} {82}},\ \bibinfo {pages} {031139} (\bibinfo
  {year} {2010})}\BibitemShut {NoStop}%
\bibitem [{\citenamefont {Zheng}(1998)}]{zhe98}%
  \BibitemOpen
  \bibfield  {author} {\bibinfo {author} {\bibfnamefont {B.}~\bibnamefont
  {Zheng}},\ }\href@noop {} {\bibfield  {journal} {\bibinfo  {journal} {Int. J.
  Mod. Phys. B}\ }\textbf {\bibinfo {volume} {12}},\ \bibinfo {pages} {1419}
  (\bibinfo {year} {1998})}\BibitemShut {NoStop}%
\bibitem [{\citenamefont {{N.J. Zhou and B. Zheng}}(2007)}]{zho07}%
  \BibitemOpen
  \bibfield  {author} {\bibinfo {author} {\bibnamefont {{N.J. Zhou and B.
  Zheng}}},\ }\href@noop {} {\bibfield  {journal} {\bibinfo  {journal}
  {Europhys. Lett.}\ }\textbf {\bibinfo {volume} {78}},\ \bibinfo {pages}
  {56001} (\bibinfo {year} {2007})}\BibitemShut {NoStop}%
\bibitem [{\citenamefont {Zhou}\ and\ \citenamefont {Zheng}(2008)}]{zho08}%
  \BibitemOpen
  \bibfield  {author} {\bibinfo {author} {\bibfnamefont {N.~J.}\ \bibnamefont
  {Zhou}}\ and\ \bibinfo {author} {\bibfnamefont {B.}~\bibnamefont {Zheng}},\
  }\href@noop {} {\bibfield  {journal} {\bibinfo  {journal} {Phys. Rev. E}\
  }\textbf {\bibinfo {volume} {77}},\ \bibinfo {pages} {051104} (\bibinfo
  {year} {2008})}\BibitemShut {NoStop}%
\bibitem [{\citenamefont {Fedorenko}\ and\ \citenamefont
  {Stepanow}(2005)}]{fed05}%
  \BibitemOpen
  \bibfield  {author} {\bibinfo {author} {\bibfnamefont {A.~A.}\ \bibnamefont
  {Fedorenko}}\ and\ \bibinfo {author} {\bibfnamefont {S.}~\bibnamefont
  {Stepanow}},\ }\href@noop {} {\bibfield  {journal} {\bibinfo  {journal}
  {Phase Transitions}\ }\textbf {\bibinfo {volume} {78}},\ \bibinfo {pages}
  {817} (\bibinfo {year} {2005})}\BibitemShut {NoStop}%
\bibitem [{\citenamefont {Zhou}\ \emph {et~al.}(2013)\citenamefont {Zhou},
  \citenamefont {Zheng},\ and\ \citenamefont {Dai}}]{zho13}%
  \BibitemOpen
  \bibfield  {author} {\bibinfo {author} {\bibfnamefont {N.~J.}\ \bibnamefont
  {Zhou}}, \bibinfo {author} {\bibfnamefont {B.}~\bibnamefont {Zheng}}, \ and\
  \bibinfo {author} {\bibfnamefont {J.~H.}\ \bibnamefont {Dai}},\ }\href@noop
  {} {\bibfield  {journal} {\bibinfo  {journal} {Phys. Rev. E}\ }\textbf
  {\bibinfo {volume} {87}},\ \bibinfo {pages} {022113} (\bibinfo {year}
  {2013})}\BibitemShut {NoStop}%
\bibitem [{\citenamefont {Jost}\ and\ \citenamefont {Usadel}(1996)}]{jos96}%
  \BibitemOpen
  \bibfield  {author} {\bibinfo {author} {\bibfnamefont {M.}~\bibnamefont
  {Jost}}\ and\ \bibinfo {author} {\bibfnamefont {K.~D.}\ \bibnamefont
  {Usadel}},\ }\href@noop {} {\bibfield  {journal} {\bibinfo  {journal} {Phys.
  Rev. B}\ }\textbf {\bibinfo {volume} {54}},\ \bibinfo {pages} {9314}
  (\bibinfo {year} {1996})}\BibitemShut {NoStop}%
\bibitem [{\citenamefont {Dong}\ \emph
  {et~al.}(2012{\natexlab{b}})\citenamefont {Dong}, \citenamefont {Zheng},\
  and\ \citenamefont {Zhou}}]{don12}%
  \BibitemOpen
  \bibfield  {author} {\bibinfo {author} {\bibfnamefont {R.~H.}\ \bibnamefont
  {Dong}}, \bibinfo {author} {\bibfnamefont {B.}~\bibnamefont {Zheng}}, \ and\
  \bibinfo {author} {\bibfnamefont {N.~J.}\ \bibnamefont {Zhou}},\ }\href@noop
  {} {\bibfield  {journal} {\bibinfo  {journal} {Europhys. Lett.}\ }\textbf
  {\bibinfo {volume} {99}},\ \bibinfo {pages} {56001} (\bibinfo {year}
  {2012}{\natexlab{b}})}\BibitemShut {NoStop}%
\bibitem [{\citenamefont {Yang}\ and\ \citenamefont {Lu}(1995)}]{yan95}%
  \BibitemOpen
  \bibfield  {author} {\bibinfo {author} {\bibfnamefont {H.~N.}\ \bibnamefont
  {Yang}}\ and\ \bibinfo {author} {\bibfnamefont {T.~M.}\ \bibnamefont {Lu}},\
  }\href@noop {} {\bibfield  {journal} {\bibinfo  {journal} {Phys. Rev. B}\
  }\textbf {\bibinfo {volume} {51}},\ \bibinfo {pages} {2479} (\bibinfo {year}
  {1995})}\BibitemShut {NoStop}%
\bibitem [{\citenamefont {Sarma}\ \emph {et~al.}(1996)\citenamefont {Sarma},
  \citenamefont {Lanczycki}, \citenamefont {Kotlyar},\ and\ \citenamefont
  {Ghaisas}}]{sar96}%
  \BibitemOpen
  \bibfield  {author} {\bibinfo {author} {\bibfnamefont {S.~D.}\ \bibnamefont
  {Sarma}}, \bibinfo {author} {\bibfnamefont {C.~J.}\ \bibnamefont
  {Lanczycki}}, \bibinfo {author} {\bibfnamefont {R.}~\bibnamefont {Kotlyar}},
  \ and\ \bibinfo {author} {\bibfnamefont {S.~V.}\ \bibnamefont {Ghaisas}},\
  }\href@noop {} {\bibfield  {journal} {\bibinfo  {journal} {Phys. Rev. E}\
  }\textbf {\bibinfo {volume} {53}},\ \bibinfo {pages} {359} (\bibinfo {year}
  {1996})}\BibitemShut {NoStop}%
\bibitem [{\citenamefont {Fedorenko}\ \emph {et~al.}(2004)\citenamefont
  {Fedorenko}, \citenamefont {Mueller},\ and\ \citenamefont
  {Stepanow}}]{fed04}%
  \BibitemOpen
  \bibfield  {author} {\bibinfo {author} {\bibfnamefont {A.~A.}\ \bibnamefont
  {Fedorenko}}, \bibinfo {author} {\bibfnamefont {V.}~\bibnamefont {Mueller}},
  \ and\ \bibinfo {author} {\bibfnamefont {S.}~\bibnamefont {Stepanow}},\
  }\href@noop {} {\bibfield  {journal} {\bibinfo  {journal} {Phys. Rev. B}\
  }\textbf {\bibinfo {volume} {70}},\ \bibinfo {pages} {224104} (\bibinfo
  {year} {2004})}\BibitemShut {NoStop}%
\bibitem [{\citenamefont {Agoritsas}\ \emph {et~al.}(2013)\citenamefont
  {Agoritsas}, \citenamefont {Lecomte},\ and\ \citenamefont
  {Giamarchi}}]{ago13}%
  \BibitemOpen
  \bibfield  {author} {\bibinfo {author} {\bibfnamefont {E.}~\bibnamefont
  {Agoritsas}}, \bibinfo {author} {\bibfnamefont {V.}~\bibnamefont {Lecomte}},
  \ and\ \bibinfo {author} {\bibfnamefont {T.}~\bibnamefont {Giamarchi}},\
  }\href@noop {} {\bibfield  {journal} {\bibinfo  {journal} {Phys. Rev. E}\
  }\textbf {\bibinfo {volume} {87}},\ \bibinfo {pages} {062405} (\bibinfo
  {year} {2013})}\BibitemShut {NoStop}%
\bibitem [{\citenamefont {Iguain}\ \emph {et~al.}(2009)\citenamefont {Iguain},
  \citenamefont {Bustingorry}, \citenamefont {Kolton},\ and\ \citenamefont
  {Cugliandolo}}]{igu09}%
  \BibitemOpen
  \bibfield  {author} {\bibinfo {author} {\bibfnamefont {J.~L.}\ \bibnamefont
  {Iguain}}, \bibinfo {author} {\bibfnamefont {S.}~\bibnamefont {Bustingorry}},
  \bibinfo {author} {\bibfnamefont {A.~B.}\ \bibnamefont {Kolton}}, \ and\
  \bibinfo {author} {\bibfnamefont {L.~F.}\ \bibnamefont {Cugliandolo}},\
  }\href@noop {} {\bibfield  {journal} {\bibinfo  {journal} {Phys. Rev. B}\
  }\textbf {\bibinfo {volume} {80}},\ \bibinfo {pages} {094201} (\bibinfo
  {year} {2009})}\BibitemShut {NoStop}%
\bibitem [{\citenamefont {Nattermann}\ \emph {et~al.}(1990)\citenamefont
  {Nattermann}, \citenamefont {Shapir},\ and\ \citenamefont {Vilfan}}]{nat90}%
  \BibitemOpen
  \bibfield  {author} {\bibinfo {author} {\bibfnamefont {T.}~\bibnamefont
  {Nattermann}}, \bibinfo {author} {\bibfnamefont {Y.}~\bibnamefont {Shapir}},
  \ and\ \bibinfo {author} {\bibfnamefont {I.}~\bibnamefont {Vilfan}},\
  }\href@noop {} {\bibfield  {journal} {\bibinfo  {journal} {Phys. Rev. B}\
  }\textbf {\bibinfo {volume} {42}},\ \bibinfo {pages} {8577} (\bibinfo {year}
  {1990})}\BibitemShut {NoStop}%
\bibitem [{\citenamefont {Tanguy}\ and\ \citenamefont
  {Vettorel}(2004)}]{tan04}%
  \BibitemOpen
  \bibfield  {author} {\bibinfo {author} {\bibfnamefont {A.}~\bibnamefont
  {Tanguy}}\ and\ \bibinfo {author} {\bibfnamefont {T.}~\bibnamefont
  {Vettorel}},\ }\href@noop {} {\bibfield  {journal} {\bibinfo  {journal} {Eur.
  Phys. J. B}\ }\textbf {\bibinfo {volume} {38}},\ \bibinfo {pages} {71}
  (\bibinfo {year} {2004})}\BibitemShut {NoStop}%
\bibitem [{\citenamefont {Nogawa}\ \emph {et~al.}(2008)\citenamefont {Nogawa},
  \citenamefont {Nemoto},\ and\ \citenamefont {Yoshino}}]{nog08}%
  \BibitemOpen
  \bibfield  {author} {\bibinfo {author} {\bibfnamefont {T.}~\bibnamefont
  {Nogawa}}, \bibinfo {author} {\bibfnamefont {K.}~\bibnamefont {Nemoto}}, \
  and\ \bibinfo {author} {\bibfnamefont {H.}~\bibnamefont {Yoshino}},\
  }\href@noop {} {\bibfield  {journal} {\bibinfo  {journal} {Phys. Rev. B}\
  }\textbf {\bibinfo {volume} {77}},\ \bibinfo {pages} {064204} (\bibinfo
  {year} {2008})}\BibitemShut {NoStop}%
\bibitem [{\citenamefont {Corberi}\ \emph {et~al.}(2011)\citenamefont
  {Corberi}, \citenamefont {Cugliandolo},\ and\ \citenamefont
  {Yoshino}}]{cor11}%
  \BibitemOpen
  \bibfield  {author} {\bibinfo {author} {\bibfnamefont {F.}~\bibnamefont
  {Corberi}}, \bibinfo {author} {\bibfnamefont {L.~F.}\ \bibnamefont
  {Cugliandolo}}, \ and\ \bibinfo {author} {\bibfnamefont {H.}~\bibnamefont
  {Yoshino}},\ }in\ \href@noop {} {\emph {\bibinfo {booktitle} {Dynamical
  Heterogeneities in Glasses, Colloids, and Granular Media}}}\ (\bibinfo {year}
  {2011})\ p.\ \bibinfo {pages} {370}\BibitemShut {NoStop}%
\bibitem [{\citenamefont {Ramasco}\ \emph {et~al.}(2000)\citenamefont
  {Ramasco}, \citenamefont {L\'opez},\ and\ \citenamefont
  {Rodr\'iguez}}]{ram00}%
  \BibitemOpen
  \bibfield  {author} {\bibinfo {author} {\bibfnamefont {J.~J.}\ \bibnamefont
  {Ramasco}}, \bibinfo {author} {\bibfnamefont {J.~M.}\ \bibnamefont
  {L\'opez}}, \ and\ \bibinfo {author} {\bibfnamefont {M.~A.}\ \bibnamefont
  {Rodr\'iguez}},\ }\href@noop {} {\bibfield  {journal} {\bibinfo  {journal}
  {Phys. Rev. Lett.}\ }\textbf {\bibinfo {volume} {84}},\ \bibinfo {pages}
  {2199} (\bibinfo {year} {2000})}\BibitemShut {NoStop}%
\bibitem [{\citenamefont {Lemerle}\ \emph {et~al.}(1998)\citenamefont
  {Lemerle}, \citenamefont {Ferr\'{e}}, \citenamefont {Chappert}, \citenamefont
  {Mathet}, \citenamefont {Giamarchi},\ and\ \citenamefont {Doussal}}]{lem98}%
  \BibitemOpen
  \bibfield  {author} {\bibinfo {author} {\bibfnamefont {S.}~\bibnamefont
  {Lemerle}}, \bibinfo {author} {\bibfnamefont {J.}~\bibnamefont {Ferr\'{e}}},
  \bibinfo {author} {\bibfnamefont {C.}~\bibnamefont {Chappert}}, \bibinfo
  {author} {\bibfnamefont {V.}~\bibnamefont {Mathet}}, \bibinfo {author}
  {\bibfnamefont {T.}~\bibnamefont {Giamarchi}}, \ and\ \bibinfo {author}
  {\bibfnamefont {P.~L.}\ \bibnamefont {Doussal}},\ }\href@noop {} {\bibfield
  {journal} {\bibinfo  {journal} {Phys. Rev. Lett.}\ }\textbf {\bibinfo
  {volume} {80}},\ \bibinfo {pages} {849} (\bibinfo {year} {1998})}\BibitemShut
  {NoStop}%
\bibitem [{\citenamefont {Lee}\ \emph {et~al.}(2009)\citenamefont {Lee},
  \citenamefont {Lee}, \citenamefont {Cho}, \citenamefont {Seo}, \citenamefont
  {Kim},\ and\ \citenamefont {Choe}}]{lee09}%
  \BibitemOpen
  \bibfield  {author} {\bibinfo {author} {\bibfnamefont {K.~S.}\ \bibnamefont
  {Lee}}, \bibinfo {author} {\bibfnamefont {C.~W.}\ \bibnamefont {Lee}},
  \bibinfo {author} {\bibfnamefont {Y.~J.}\ \bibnamefont {Cho}}, \bibinfo
  {author} {\bibfnamefont {S.}~\bibnamefont {Seo}}, \bibinfo {author}
  {\bibfnamefont {D.~H.}\ \bibnamefont {Kim}}, \ and\ \bibinfo {author}
  {\bibfnamefont {S.~B.}\ \bibnamefont {Choe}},\ }\href@noop {} {\bibfield
  {journal} {\bibinfo  {journal} {IEEE. Trans. Magn.}\ }\textbf {\bibinfo
  {volume} {45}},\ \bibinfo {pages} {2548} (\bibinfo {year}
  {2009})}\BibitemShut {NoStop}%
\end{thebibliography}%
\bibliographystyle{apsrev4-1}

\begin{table}[h]\centering
\caption{Scaling exponents of the creep dynamics at different
temperatures $T$. The scaling relation $\psi \beta  / \zeta_{loc} = 0.45(3)$ holds within error bars.
The roughness exponents $\zeta=0.53(1)$ and $0.68(1)$ are measured in the small-$D\xi(t)$ regime (central columns) and
large-$D\xi(t)$ regime (right columns), respectively, in good agreement with the local ones $\zeta_{loc}$.}
\begin{tabular}[t]{l| c c c c| c c}
\hline
\hline   T       &  \quad$0.025$    &  \quad$0.05$      &  \quad $0.1$        & \quad$0.2$ \quad        &  \quad$0.33$      & \quad$0.66$        \\
\hline
        $\beta$  & \quad$0.23(1)$   &  \quad$0.23(1)$   &  \quad$0.26(1)$     & \quad$0.38(1)$ \quad    &  \quad$0.55(2)$   & \quad$0.52(2)$      \\
        $1/\psi$ & \quad$0.98(1)$   &  \quad$0.98(1)$   &  \quad$1.16(1)$     & \quad$1.73(2)$ \quad    &  \quad$1.84(2)$   & \quad$1.75(3)$  \\
  $\zeta_{loc}$  & \quad$0.50(1)$   &  \quad$0.51(1)$   &  \quad$0.50(1)$     & \quad$0.51(1)$ \quad    &  \quad$0.65(1)$   & \quad$0.69(1)$  \\
        \hline
$ \psi \beta / \zeta_{loc} $
                 & \quad$0.47(3)$   &  \quad $0.46(3)$  &  \quad $0.45(3)$    & \quad $0.43(3)$ \quad   &  \quad$0.46(3)$   & \quad$0.43(3)$ \\
        $\zeta$  &    \multicolumn{4}{c|}{\quad$0.53(1)$ \quad}&  \multicolumn{2}{c}{\quad$0.68(1)$} \\
\hline \hline
\end{tabular}
\label{t1}
\end{table}

\begin{table}[h]\centering
\caption{Scaling exponents in the large-$D\xi(t)$ regime at different
disorders $\Delta$, driving fields $H_0$ and frequencies $f$. As the creep correlation length $D\xi(t)$ grows,
a significant deviation of the roughness exponent $\zeta$ from $\zeta_{loc} \approx 2/3$ is observed in the right columns.}
\begin{tabular}[t]{l c c c c  c}
\hline
\hline  $\Delta$  &  \quad$1.5$       & \quad $1.0$      &  \quad $0.7$ \quad     &  \quad $0.5$     & \quad $0.1$ \\
        $1/\psi$  &  \quad$1.84(1)$   & \quad $1.89(2)$  &  \quad $1.82(2)$ \quad &  \quad $1.85(3)$ & \quad $3.07(3)$  \\
   $\zeta_{loc}$  &  \quad$0.65(1)$   & \quad $0.66(1)$  &  \quad $0.66(1)$ \quad &  \quad $0.67(1)$ & \quad $0.66(1)$ \\
        $\zeta$   &  \multicolumn{3}{c}{$\zeta=0.68(1)$} & \multicolumn{2}{c}{$\zeta=0.78(1)$} \\ \hline

        $H_0$     & \quad$0.01$      & \quad $0.05$     & \quad $0.1$ \quad     &  \quad $0.2$     &  \quad $0.5$    \\
        $1/\psi$  &  \quad $1.84(2)$ & \quad $1.86(2)$  & \quad $1.90(2)$ \quad  &  \quad $2.11(2)$ &  \quad $2.62(2)$  \\
   $\zeta_{loc}$  &  \quad$0.65(1)$  & \quad $0.65(1)$  &  \quad $0.66(1)$ \quad &  \quad $0.68(1)$ & \quad $0.66(1)$ \\
        $\zeta$   &  \multicolumn{3}{c}{$\zeta=0.68(1)$} & \multicolumn{2}{c}{$\zeta=0.94(2)$} \\ \hline

        $f$ (Hz)  & \quad$10^{-4}$   & \quad $ 10^{-3}$ &  \quad $10^{-2}$ \quad & \quad $10^{-1}$ & \quad $10^{0}$\\
        $1/\psi$  &  \quad $1.90(2)$ & \quad $1.90(3)$  &  \quad $1.88(3)$ \quad& \quad $1.87(3)$ &  \quad  1.84(3)\\
    $\zeta_{loc}$ &  \quad$0.66(1)$  & \quad $0.66(1)$  &  \quad $0.65(1)$ \quad &  \quad $0.65(1)$ & \quad $0.65(1)$ \\
        $\zeta$  &  \multicolumn{3}{c}{$\zeta=0.69(1)$} & \multicolumn{2}{c}{$\zeta=0.84(1)$} \\ \hline \hline
\end{tabular}
\label{t2}
\end{table}

\clearpage

\end{document}